\documentclass[11pt,a4paper]{article}

\usepackage{jheppub}

\usepackage{graphicx} 

\usepackage{amsmath}
\usepackage{amscd}
\usepackage{amsfonts}
\usepackage{amssymb}
\usepackage{mathrsfs}
\usepackage{fancybox} 
\usepackage{enumerate}
\usepackage[boxsize=0.8ex]{ytableau} 
\usepackage{youngtab}
\usepackage{bbm} 
\usepackage{graphicx}
\usepackage{slashed}
\usepackage{color}
 \usepackage{ulem}
\usepackage{tikz}

\usepackage{comment}

\def\={\stackrel{\bullet}{=}}

\def\({\left(}
\def\){\right)}
\def\[{\left[}
\def\]{\right]}

\def \be {\begin{equation}}
\def \ee {\end{equation}}
\def \beqa {\begin{eqnarray}}
\def \eeqa {\end{eqnarray}}
\def \beal#1 {\begin{align}#1\end{align}}
\def \bes#1 {\begin{equation}\begin{split}#1\end{split}\end{equation}}

\def\x't{(\boldsymbol{x'},t)}

\def\3tensor#1#2#3#4{#1^{#2\;#4}_{\;\;#3}}

\def\z#1{\mathbb{Z}_{#1}} 

\def\i{\mathrm{i}}

\begin{document}

\vspace{-2cm}
\begin{flushright}
\normalsize{ 
 OU-HET-1306}
\end{flushright}

\title{Taste-splitting mass and edge modes in $3+1$~D staggered fermions  }

\author[a]{Tatsuhiro Misumi,\footnote{\it misumi@phys.kindai.ac.jp}  }
\affiliation[a]{Department of Physics, Kindai University, Higashi-Osaka, Osaka 577-8502, Japan}
\author[b]{Tetsuya Onogi,\footnote{\it onogi@het.phys.sci.osaka-u.ac.jp} }
\author[b]{Tatsuya Yamaoka\footnote{\it t\_yamaoka@het.phys.sci.osaka-u.ac.jp}}     
\affiliation[b]{Department of Physics, The University of Osaka, Toyonaka, Osaka 560-0043, Japan}

\abstract{
We investigate the symmetry structure of the $3+1$~D staggered fermion Hamiltonian and its implications for anomalies. 
Since the spin and flavor degrees of freedom of Dirac fermions are distributed over the lattice, in addition to the standard on-site mass term, the staggered fermion system also admits one-, two-, and three-link bilinear terms within a unit cube as local, charge conserving mass terms with different spin and flavor dependence.
We identify the spin flavor structures of all those bilinear mass terms and determine the symmetries preserved by each of them.
Among them, one of the one-link mass terms preserves a larger
residual symmetry associated with conserved charges that generate the Onsager algebra.
Motivated by this structure, we consider a kink profile of the one-link mass and analyze the resulting domain-wall system.
In the low-energy limit, the $3+1$~D bulk becomes gapped, while two-flavor massless Dirac fermions appear as localized modes on the $2+1$~D domain wall. 
We show that the bulk conserved charges act on the wall as generators of a flavor $\mathrm{SU}(2)$ symmetry, and that no symmetric mass gap is allowed for the boundary theory when this $\mathrm{SU}(2)$ symmetry and space reflection symmetry are both imposed. 
This realizes the parity anomaly of the boundary theory and shows that the boundary flavor symmetry and anomaly descend from the ultraviolet staggered-fermion Hamiltonian rather than emerging only in the infrared.  
}

\date{today}

\maketitle

\section{Introduction}
\label{sec:Introduction}

Lattice field theory provides a powerful framework for studying the nonperturbative dynamics of quantum field theories, and has been widely applied in both high-energy physics and condensed matter physics. 
However, according to the Nielsen--Ninomiya no-go theorem~\cite{Nielsen:1980rz,Nielsen:1981hk}, a naive lattice discretization of fermions inevitably introduces additional low-energy degrees of freedom, known as doublers, in the infrared (IR) effective theory. 
For instance, in a $3+1$~D Hamiltonian system, $2^3 = 8$ fermionic doublers appear. 
This phenomenon is referred to as the doubling problem.

A well-known approach to resolving this problem is the Wilson fermion formulation~\cite{Wilson:1975id}. 
By introducing a momentum-dependent mass term into the Hamiltonian, one can remove seven of the doublers. 
However, this procedure explicitly breaks chiral symmetry.

Alternatively, one may take advantage of the doubling phenomenon by reinterpreting the doublers as flavor degrees of freedom and incorporating them into the theory~\cite{Kogut:1974ag,Susskind:1976jm}. 
Such systems are known as staggered fermion theories. 
In this formulation, the Hamiltonian is described by one-component complex fermions obtained by staggering the spin and flavor degrees of freedom of Dirac fermions across the spatial lattice.

In recent years, the symmetry structure and anomalies of staggered fermion Hamiltonian systems have been actively investigated~\cite{Li:2024dpq,Chatterjee:2024gje,Catterall:2025vrx,Yamaoka:2025sdm,Onogi:2025xir,Aoki:2025vtp,Seiberg:2026icc}. \footnote{Symmetries and anomalies in Wilson fermion systems were also discussed by~\cite{Misumi:2019jrt}.}
Because the spin and flavor degrees of freedom are distributed over the lattice in a staggered manner, the theory exhibits a particularly rich symmetry structure. 
Crystalline symmetries of the lattice correspond, in the continuum limit, to internal symmetries that simultaneously transform the spin and flavor components of Dirac fermions. 
Consequently, anomaly matching between the ultraviolet (UV) lattice theory and the infrared (IR) continuum theory becomes highly nontrivial.

A particularly interesting recent development concerns conserved charges of the staggered fermion Hamiltonian that generate the Onsager algebra~\cite{Onsager:1943jn,Chatterjee:2024gje}.
Remarkably, when the spatial dimension is odd (such as one or three), these conserved charges generate chiral flavor symmetries, or their subgroups, in the continuum limit~\cite{Gioia:2025bhl,Yamaoka:2025sdm,Onogi:2025xir,Aoki:2025vtp}.
In contrast, in two spatial dimensions, the corresponding conserved charges generate an $\mathrm{SU}(2)$ flavor symmetry in the continuum theory~\cite{Pace:2025rfu}. 
For example, in the honeycomb lattice system this symmetry is identified with the so-called valley symmetry. 
Furthermore, in each dimension these conserved charges correctly reproduce the anomaly structure of the IR theory.

These observations lead to an important conceptual insight: part of the symmetry and anomaly structure that was previously thought to emerge only in the infrared actually originates from symmetries and anomalies already present in the ultraviolet lattice theory. 
Therefore, understanding the symmetries and 't~Hooft
anomalies of lattice field theories plays an essential role in elucidating the structure of the corresponding continuum infrared theories.


In this paper, we revisit the $3+1$~D staggered fermion system in the Hamiltonian formalism and address two main questions. 
First, we classify all Hermitian, particle-number-preserving bilinear mass terms local within a unit cube that gap the continuum Dirac fermions and clarify the symmetries preserved by each of them.
Second, we investigate the symmetry structure and anomalies of the $3+1$~D staggered fermion Hamiltonian in the presence of a $2+1$~D boundary (domain wall), with particular emphasis on the physical role of the conserved charges that generate the Onsager algebra.

The first problem is motivated from two perspectives. 
One motivation is that the classification of mass terms and their symmetries provides a useful diagnostic for identifying 't~Hooft anomalies. 't~Hooft anomalies can be interpreted as obstructions to realizing a symmetric, trivially gapped phase~\cite{Lieb:1961fr,Affleck:1986pq,Oshikawa2000TopologicalAT,Hastings:2003zx,Chang:2018iay,Wen:2018zux,Thorngren:2019iar}. 
In particular, the absence of a mass term that preserves a given symmetry often signals the presence of a corresponding 't~Hooft anomaly. 
Another motivation is more practical. 
If one aims to perform numerical studies of theories such as QCD using the staggered fermion Hamiltonian formalism in the future, it is important to understand how meson operators are constructed and which symmetries they respect.\footnote{In the Lagrangian formalism, these issues have been systematically studied, for example, in Refs.~\cite{Golterman:1984cy,Golterman:1985dz,Kilcup:1986dg}.}

By introducing suitable mass terms, it is possible to gap out a subset of flavor fermions and thereby obtain an effective theory with fewer fermionic degrees of freedom. 
This idea underlies the species-splitting (flavored) mass for the minimal-doubling models~\cite{Karsten:1981gd,Wilczek:1987kw,Creutz:2007af,Borici:2007kz,Creutz:2010qm,Creutz:2010cz,Creutz:2010bm,Wan2010TopologicalSA} and has also been actively studied as a possible route toward realizing chiral gauge theories on the lattice~\cite{Gioia:2025bhl,Misumi:2025yjf}. 
Inspired by this idea, as well as by the construction of domain-wall fermions~\cite{Kaplan:1992bt,Shamir:1993zy,Hoelbling:2016qfv,Hoelbling:2016dug}, we introduce into the $3+1$~D staggered fermion Hamiltonian a mass term classified in response to the first question with a kink-type spatial profile.

In the low-energy limit, the bulk $3+1$~D system becomes gapped, while two-flavor massless Dirac fermions are expected to appear as zero modes localized on the $2+1$~D domain wall.
If this scenario is realized, the resulting edge theory is known to exhibit the parity anomaly associated with the flavor $\mathrm{SU}(2)$ symmetry~\cite{Redlich:1983dv,Redlich:1983kn,Seiberg:2016rsg}.
This naturally raises the following question: is this symmetry and its associated anomaly emergent in the infrared, or does it originate from structures already present in the ultraviolet lattice theory?

With the  ``normal definition'' of staggered phase,
we show that compared to other mass terms, the one-link mass in the $x$-direction preserves the largest residual symmetry, including the symmetry associated with the conserved charges that generate the Onsager algebra.\footnote{It should be noted that this result strongly depends on the choice of the staggered phases; see footnote~\ref{footnote:staggered-phase} and Sec.~\ref{subsec:one-link-mass}.}
Motivated by this structure, we consider a kink profile of the one-link mass and analyze the resulting domain-wall system. 
In this system, remarkably, the conserved charges in the $3+1$~D bulk theory that generate the Onsager algebra---whose continuum limit realize chiral flavor subgroups $\mathrm{U}(1)_{F_i} \subset \mathrm{SU}(2)\times \mathrm{SU}(2)\times \mathrm{U}(1)_A$---act as generators of a flavor $\mathrm{SU}(2)$ symmetry in the $2+1$~D edge theory. 
Furthermore, we demonstrate that the edge theory cannot be gapped while preserving both this flavor $\mathrm{SU}(2)$ symmetry and space reflection symmetry. 
This result implies that, in our model, the flavor $\mathrm{SU}(2)$ symmetry and its associated parity anomaly are not emergent phenomena, but instead originate from the symmetry structure already present in the $3+1$~D bulk theory.

This paper is organized as follows. 
In Sec.~\ref{sec:Hamiltonian-lattice}, we briefly review the Hamiltonian formulation of a single staggered fermion. 
In Sec.~\ref{sec:symmetries}, we discuss the discrete spacetime and internal symmetries of the staggered fermion system. 
Sec.~\ref{sec:classification-mass} classifies all possible bilinear Dirac mass terms within this framework. 
Building on these discussions, in Sec.~\ref{sec:mass-splitting} we explore the $2+1$~D boundary theory of the $3+1$~D staggered fermion system by introducing a kink profile of the $x$-directed one-link mass. 
Section~\ref{sec:Conclusion} is devoted to our conclusions. 
Finally, a parallel discussion of the symmetries and mass classification within the Lagrangian formalism is provided in Appendix~\ref{sec:appendix_staggered_wilson}, and in Appendix~\ref{sec:app-extraEvidence} we present additional evidence that the flavor $\mathrm{SU}(2)$ symmetry on the edge is emanant: the bulk charges forming $\mathrm{Ons}\,3$ generate the full $\mathrm{SU}(2)_L \times \mathrm{SU}(2)_R$ symmetry, whose explicit breaking to $\mathrm{Ons}\,2$ leaves precisely this edge symmetry.

\section{Staggered fermion Hamiltonian on the lattice}
\label{sec:Hamiltonian-lattice}
\subsection{Setup}
\label{subsec:setup}

The Hamiltonian of the $3+1$-dimensional staggered fermion is given by
\begin{align}
    \label{eq:Hamiltonian-staggerd}
    H = \sum_{\boldsymbol{r}}^{L}\sum_{i=1}^{3} \frac{\mathrm{i}}{2} \eta_i (\boldsymbol{r}) 
    \left( \chi (\boldsymbol{r})^\dagger \chi (\boldsymbol{r}+ \hat{i}) 
    + \chi (\boldsymbol{r}) \chi^\dagger (\boldsymbol{r}+ \hat{i}) \right)  \, ,
\end{align}
where the spatial coordinate vector is denoted by $\boldsymbol{r}=(r_1, r_2, r_3) = (x,y,z)$.
The unit vector in the $i$-direction is denoted as $\hat{i}$ ($i = x, y, z$) and
the staggered phase $\eta_i(\boldsymbol{r})$ is given by~\footnote{
In this paper, we employ the staggered phase~\eqref{eq:staggered-phase}, which generates a $\pi$ flux on each plane within a unit cell. As discussed in Sec.~\ref{subsec:two-flavorDirac}, this $\pi$ flux is essential to realize eight zero-modes, corresponding to two-flavor four-component Dirac fermions. This choice also naturally yields the standard finite-difference kinetic terms in the Hamiltonian~\eqref{eq:Hamiltonian-staggerd}. Up to the arbitrariness of unitary transformations associated with sublattice redefinitions, we refer to this specific class of phases as the ``normal definition.''
\label{footnote:staggered-phase}
}
\begin{align}
    \label{eq:staggered-phase}
    \eta_i(\boldsymbol{r}) = (-1)^{r_1 + \dots + r_{i-1}}  \, .
\end{align}
The complex fermion $\chi$ can be decomposed into two Majorana fermions $a$ and $b$, satisfying
\begin{align}
    a (\boldsymbol{r}) =  a^\dagger (\boldsymbol{r}), 
    \qquad 
    b (\boldsymbol{r}) = b^\dagger (\boldsymbol{r}), 
    \qquad 
    \{a (\boldsymbol{r}),a^\dagger (\boldsymbol{r}')\} 
    = \{b (\boldsymbol{r}),b^\dagger (\boldsymbol{r}')\} 
    = 2 \delta_{\boldsymbol{r}, \boldsymbol{r}'} \, ,
\end{align}
through
\begin{align}
    \chi(\boldsymbol{r}) = \frac{1}{2} \left( a (\boldsymbol{r}) + \i b (\boldsymbol{r}) \right) \, .
\end{align}
Substituting this expression into Eq.~\eqref{eq:Hamiltonian-staggerd}, the Hamiltonian can be rewritten as
\begin{align}
    \label{eq:Majorana-Hamiltonian}
    H &= H_a + H_b = \sum_{\boldsymbol{r}}\sum_{i=1}^{3} \frac{\i}{4} \eta_i(\boldsymbol{r}) 
      \left( a({\boldsymbol{r}}) a({\boldsymbol{r}+\hat{i}}) 
      + b({\boldsymbol{r}}) b({\boldsymbol{r}+\hat{i}}) \right) \, .
\end{align}

As drawn in Figure~\ref{fig:cube_sublattices}, the lattice is composed of four sublattices within each unit cube, labeled as
\begin{align}
    \label{eq:sublattices-even} 
     (A_{\boldsymbol{r}_e}, \, B_{\boldsymbol{r}_e}, \, C_{\boldsymbol{r}_e}, \, D_{\boldsymbol{r}_e}) := (\chi({\boldsymbol{r}_{A_e}}), \chi({\boldsymbol{r}_{B_e}}), \chi({\boldsymbol{r}_{C_e}}), \chi({\boldsymbol{r}_{D_e}})) \, , 
\end{align}
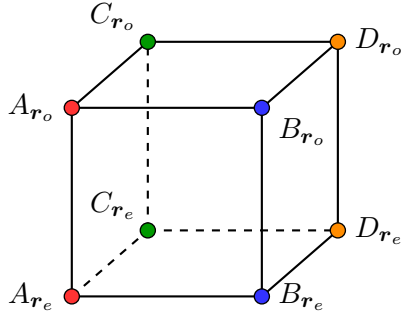
\begin{figure}[htbp]
    \centering
    \begin{tikzpicture}[
        scale=2.5,
        x={(1cm, 0cm)}, y={(0.4cm, 0.35cm)}, z={(0cm, 1cm)},
        vertex/.style={circle, inner sep=2pt, draw=black, line width=0.5pt},
        nodeA/.style={vertex, fill=red!80},
        nodeB/.style={vertex, fill=blue!80},
        nodeC/.style={vertex, fill=green!60!black},
        nodeD/.style={vertex, fill=orange!90!yellow},
        edge/.style={thick, black},
        hidden edge/.style={thick, dashed, black}
    ]

    
    \coordinate (A_re) at (0, 0, 0);
    \coordinate (B_re) at (1, 0, 0);
    \coordinate (C_re) at (0, 1, 0);
    \coordinate (D_re) at (1, 1, 0);
    
    \coordinate (A_ro) at (0, 0, 1);
    \coordinate (B_ro) at (1, 0, 1);
    \coordinate (C_ro) at (0, 1, 1);
    \coordinate (D_ro) at (1, 1, 1);

    \draw[hidden edge] (A_re) -- (C_re);
    \draw[hidden edge] (C_re) -- (D_re);
    \draw[hidden edge] (C_re) -- (C_ro);

    \draw[edge] (A_re) -- (B_re);
    \draw[edge] (B_re) -- (D_re);
    \draw[edge] (A_ro) -- (B_ro) -- (D_ro) -- (C_ro) -- cycle;
    \draw[edge] (A_re) -- (A_ro);
    \draw[edge] (B_re) -- (B_ro);
    \draw[edge] (D_re) -- (D_ro);

    \node[nodeC, label={[anchor=south east, inner sep=2pt]135:$C_{\boldsymbol{r}_e}$}] at (C_re) {};
    \node[nodeC, label={[anchor=south east, inner sep=2pt]135:$C_{\boldsymbol{r}_o}$}] at (C_ro) {};
    
    \node[nodeD, label={[anchor=west, inner sep=3pt]0:$D_{\boldsymbol{r}_e}$}] at (D_re) {};
    \node[nodeD, label={[anchor=west, inner sep=3pt]0:$D_{\boldsymbol{r}_o}$}] at (D_ro) {};

    \node[nodeA, label={[anchor=east, inner sep=3pt]180:$A_{\boldsymbol{r}_e}$}] at (A_re) {};
    \node[nodeA, label={[anchor=east, inner sep=3pt]180:$A_{\boldsymbol{r}_o}$}] at (A_ro) {};
    
    \node[nodeB, label={[anchor=west, inner sep=3pt]0:$B_{\boldsymbol{r}_e}$}] at (B_re) {};
    \node[nodeB, label={[anchor=north west, inner sep=3pt]-15:$B_{\boldsymbol{r}_o}$}] at (B_ro) {};

    \end{tikzpicture}
    \caption{Schematic representation of the sublattices within a unit cube.
    }
    \label{fig:cube_sublattices}
\end{figure}
where
\begin{align}
    \boldsymbol{r}_e &= \boldsymbol{r}_{A_e}=  (2n_x, 2n_y, 2n_z) \, ,  \\
    \boldsymbol{r}_{B_e} &= \boldsymbol{r}_{A_e} + \hat{x} \, , \quad
    \boldsymbol{r}_{C_e} = \boldsymbol{r}_{A_e} + \hat{y} \, , \quad
    \boldsymbol{r}_{D_e} = \boldsymbol{r}_{A_e}+ \hat{x} + \hat{y} \, ,
\end{align}
or equivalently
\begin{align}
    \label{eq:sublattices-odd}
 (A_{\boldsymbol{r}_o}, \, B_{\boldsymbol{r}_o}, \, C_{\boldsymbol{r}_o}, \, D_{\boldsymbol{r}_o}) := (\chi({\boldsymbol{r}_{A_o}}), \chi({\boldsymbol{r}_{B_o}}), \chi({\boldsymbol{r}_{C_o}}), \chi({\boldsymbol{r}_{D_o}})) \, , 
\end{align}
where
\begin{align}
    \boldsymbol{r}_o &= \boldsymbol{r}_{A_o}=  (2n_x, 2n_y, 2n_z + 1) \, , \\
    \boldsymbol{r}_{B_o} &= \boldsymbol{r}_{A_o} + \hat{x} \, , \quad
    \boldsymbol{r}_{C_o} = \boldsymbol{r}_{A_o} + \hat{y} \, , \quad
    \boldsymbol{r}_{D_o} = \boldsymbol{r}_{A_o} + \hat{x} + \hat{y} \, .
\end{align}

The fermions on each sublattice are Fourier expanded as
\begin{align}
    \label{eq:Fouriermode}
    A_{\boldsymbol{r}_e} 
    = \frac{2}{\sqrt{L^3}} \sum_{\boldsymbol{k} \in \text{BZ}} 
    e^{\i \boldsymbol{k} \cdot \boldsymbol{r}_e} \tilde{A}_{\boldsymbol{k}} \, , 
\end{align}
with analogous expressions for $B_{\boldsymbol{r}_e}$, $C_{\boldsymbol{r}_e}$, and $D_{\boldsymbol{r}_e}$.
Here, $\text{BZ}$ denotes the first Brillouin zone, and the Fourier modes satisfy
$\{ \tilde{A}_{\boldsymbol{k}} , \tilde{A}^\dagger_{\boldsymbol{k'}} \} 
= \delta_{\boldsymbol{k},\boldsymbol{k'}}$.

In the continuum limit, the low-energy modes are localized near the momenta
\begin{align}
    \label{eq:momentum-of-zeromode}
    \boldsymbol{K}_0 = (0,0,0), 
    \qquad 
    \boldsymbol{K}_1 = (0,0,\pi) \ .
\end{align}

\subsection{Two-flavor free, massless Dirac fermions}
\label{subsec:two-flavorDirac}

We now show that the Hamiltonian~\eqref{eq:Hamiltonian-staggerd} describes two flavors of free, massless Dirac fermions. 
We begin by introducing the following matrices constructed from the gamma matrices,
\begin{align}
    \label{eq:alphaI-alphaR}
    \alpha_i &= \gamma_0 \gamma_i \, , \quad \beta = \gamma_0 \, , \notag \\
    \alpha^r &= \alpha_1^{x} \alpha_2^{y} \alpha_3^{z} \, ,
\end{align}
which satisfy the algebraic relations
\begin{align}
    \{ \alpha_i , \alpha_j \} = 2 \delta_{ij} \ , \quad 
    \{ \alpha_i , \beta \} = 0 \ , \quad 
    \alpha_i^\dagger = \alpha_i \ .
\end{align}
Throughout this section, we work in the Weyl representation,
\begin{align}
    \label{eq:gamma-WeylRep}
    \gamma_0 &= \mathbbm{1}_{2\times 2} \otimes \sigma_1 
    = \begin{pmatrix}
        0 & \mathbbm{1}_{2\times 2} \\
        \mathbbm{1}_{2\times 2} & 0
    \end{pmatrix} \, , \quad \gamma_i = \sigma_i \otimes \i \sigma_2 
    = \begin{pmatrix}
        0 & \sigma_i \\
        - \sigma_i & 0
    \end{pmatrix} \, , \notag
\end{align}
which leads to
\begin{align}
    \gamma_5 &= \i \alpha_1 \alpha_2 \alpha_3 
    = -\i \gamma_0 \gamma_1 \gamma_2 \gamma_3 
    = \mathbb{I}_{2\times 2} \otimes \sigma_3 
    = \begin{pmatrix}
        \mathbbm{1}_{2\times 2} & 0 \\
        0 & - \mathbbm{1}_{2\times 2}
    \end{pmatrix} \, , \\
    \alpha_i &= - \sigma_i \otimes \sigma_3 
    = \begin{pmatrix}
        - \sigma_i & 0 \\
        0 & \sigma_i
    \end{pmatrix} \, .
\end{align}

We next define a matrix-valued fermion field constructed from the staggered fermions as follows~\cite{Catterall:2025vrx,Onogi:2025xir},
\begin{align}
    \label{eq:Matrix-Fermion}
    \Phi ( \boldsymbol{R}_e ) 
    &= N_0 \sum_{\{ \boldsymbol{b} \}} \chi(\boldsymbol{R}_e + \boldsymbol{b}) \, \alpha^{b} 
    = \begin{pmatrix}
        \psi_{1+}(\boldsymbol{R}_e) & \psi_{2+}(\boldsymbol{R}_e) & 0 & 0 \\
        0 & 0 & \psi_{3-}(\boldsymbol{R}_e) & \psi_{4-}(\boldsymbol{R}_e)
    \end{pmatrix} \, ,
\end{align}
where $\{\boldsymbol{b}\}$ denotes the set of $2^3$ vectors with components $b_i = \{0,1\}$, 
$\boldsymbol{R}_e = \boldsymbol{r}_e$, and $N_0$ is a normalization constant. 
Here, $\psi_{1,2}$ ($\psi_{3,4}$) are two-component left-handed (right-handed) Weyl fermions defined by
\begin{align}
    \psi_{1+}(\boldsymbol{R}_e) &= \begin{pmatrix}
        A_{\boldsymbol{r}_e} -  A_{\boldsymbol{r}_o} + \mathrm{i} (  D_{\boldsymbol{r}_e} -  D_{\boldsymbol{r}_o} ) \\
        -(B_{\boldsymbol{r}_e} -  B_{\boldsymbol{r}_o}) - \mathrm{i} (  C_{\boldsymbol{r}_e} -  C_{\boldsymbol{r}_o} ) 
    \end{pmatrix} \, , \\
    \psi_{2+}(\boldsymbol{R}_e) &= \begin{pmatrix}
        -(B_{\boldsymbol{r}_e} +  B_{\boldsymbol{r}_o}) + \mathrm{i} (  C_{\boldsymbol{r}_e} +  C_{\boldsymbol{r}_o} ) \\
        A_{\boldsymbol{r}_e} +  A_{\boldsymbol{r}_o} - \mathrm{i} (  D_{\boldsymbol{r}_e} +  D_{\boldsymbol{r}_o} ) 
    \end{pmatrix} \, , \\
    \psi_{3-}(\boldsymbol{R}_e) &= \begin{pmatrix}
        A_{\boldsymbol{r}_e} +  A_{\boldsymbol{r}_o} + \mathrm{i} (  D_{\boldsymbol{r}_e} +  D_{\boldsymbol{r}_o} ) \\
        B_{\boldsymbol{r}_e} +  B_{\boldsymbol{r}_o} + \mathrm{i} (  C_{\boldsymbol{r}_e} +  C_{\boldsymbol{r}_o} ) 
    \end{pmatrix} \, , \\
    \psi_{4-}(\boldsymbol{R}_e) &= \begin{pmatrix}
        B_{\boldsymbol{r}_e} -  B_{\boldsymbol{r}_o} - \mathrm{i} (  C_{\boldsymbol{r}_e} -  C_{\boldsymbol{r}_o} ) \\
        A_{\boldsymbol{r}_e} -  A_{\boldsymbol{r}_o} - \mathrm{i} (  D_{\boldsymbol{r}_e} -  D_{\boldsymbol{r}_o} ) 
    \end{pmatrix} \, .
\end{align}
In momentum space, each Weyl fermion Fourier mode $\tilde{\psi}_f$ 
is localized around the momentum $\boldsymbol{K}_0$ or $\boldsymbol{K}_1$, 
given by Eq.~\eqref{eq:momentum-of-zeromode}, in the continuum limit,
\begin{align}
    \label{eq:Fouriermode-Contlimit-psi1}
    \tilde{\psi}_{1+}(\boldsymbol{k}) &\propto \begin{pmatrix}
        \tilde{A}_{\boldsymbol{K}_1 + \boldsymbol{k}} + \i \tilde{D}_{\boldsymbol{K}_1 + \boldsymbol{k}} \\
        -\tilde{B}_{\boldsymbol{K}_1 + \boldsymbol{k}} - \i \tilde{C}_{\boldsymbol{K}_1 + \boldsymbol{k}}
    \end{pmatrix} \, , \\
    \label{eq:Fouriermode-Contlimit-psi2}
    \tilde{\psi}_{2+}(\boldsymbol{k}) &\propto \begin{pmatrix}
        -\tilde{B}_{\boldsymbol{K}_0 + \boldsymbol{k}} + \i \tilde{C}_{\boldsymbol{K}_0 + \boldsymbol{k}} \\
        \tilde{A}_{\boldsymbol{K}_0 + \boldsymbol{k}} - \i \tilde{D}_{\boldsymbol{K}_0 + \boldsymbol{k}}
    \end{pmatrix} \, , \\
    \label{eq:Fouriermode-Contlimit-psi3}
    \tilde{\psi}_{3-}(\boldsymbol{k}) &\propto \begin{pmatrix}
        \tilde{A}_{\boldsymbol{K}_0 + \boldsymbol{k}} + \i \tilde{D}_{\boldsymbol{K}_0 + \boldsymbol{k}} \\
        \tilde{B}_{\boldsymbol{K}_0 + \boldsymbol{k}} + \i \tilde{C}_{\boldsymbol{K}_0 + \boldsymbol{k}}
    \end{pmatrix} \, , \\
    \label{eq:Fouriermode-Contlimit-psi4}
    \tilde{\psi}_{4-}(\boldsymbol{k}) &\propto \begin{pmatrix}
        \tilde{B}_{\boldsymbol{K}_1 + \boldsymbol{k}} - \i \tilde{C}_{\boldsymbol{K}_1 + \boldsymbol{k}} \\
        \tilde{A}_{\boldsymbol{K}_1 + \boldsymbol{k}} - \i \tilde{D}_{\boldsymbol{K}_1 + \boldsymbol{k}}
    \end{pmatrix} \, ,
\end{align}
where $\tilde{A}_{\boldsymbol{k}}$, $\tilde{B}_{\boldsymbol{k}}$, $\tilde{C}_{\boldsymbol{k}}$, and $\tilde{D}_{\boldsymbol{k}}$ 
are the Fourier modes of the sublattices defined in Eqs.~\eqref{eq:sublattices-even} and \eqref{eq:sublattices-odd}.

We now combine the Weyl fermions $\psi_f$ into two four-component Dirac fermions,
\begin{align}
    \psi_1 (\boldsymbol{R}_e) &= 
    \begin{pmatrix} 
        \psi_{1+}(\boldsymbol{R}_e) \\ 
        \psi_{3-}(\boldsymbol{R}_e) 
    \end{pmatrix} \, , \quad 
    \psi_2 (\boldsymbol{R}_e) = 
    \begin{pmatrix} 
        \psi_{2+}(\boldsymbol{R}_e) \\ 
        \psi_{4-}(\boldsymbol{R}_e) 
    \end{pmatrix} \, ,
\end{align}
which together form a $2 \times 2$ matrix-valued fermion field,
\begin{align}
    \label{eq:2-2-matrix-fermion}
    \Psi (\boldsymbol{R}_e) 
    = \begin{pmatrix} 
        \psi_1(\boldsymbol{R}_e) & \psi_2(\boldsymbol{R}_e) 
    \end{pmatrix} 
    = \begin{pmatrix}
        \psi_{1+} & \psi_{2+} \\
        \psi_{3-} & \psi_{4-}
    \end{pmatrix} \, .
\end{align}
The Hamiltonian~\eqref{eq:Hamiltonian-staggerd} can then be written as
\begin{align}
    \label{Hamiltonian-matrix-2-flavor}
    H &= \i \sum_{\boldsymbol{R}_e} 
    \Psi^\dagger(\boldsymbol{R}_e) 
    \left[
        (\alpha_i \otimes \mathbbm{1}_{2 \times 2}) \frac{\nabla_i}{2}  
        + (\beta \gamma_5 \otimes \sigma_i^T) \frac{\nabla_i^2}{2} 
    \right] 
    \Psi(\boldsymbol{R}_e) \, ,
\end{align}
where the matrices $\alpha_i$ and $\beta \gamma_5$ act on the spin indices, while $\sigma_i^T$ acts on the flavor indices. 
The lattice difference operator $\nabla_i$ acts on $\Psi(\boldsymbol{R}_e)$ as
\begin{align}
    \nabla_i \Psi(\boldsymbol{R}_e) 
    &= \Psi(\boldsymbol{R}_e + 2\hat{i}) - \Psi(\boldsymbol{R}_e - 2\hat{i}) \notag \\
    &= \Psi(\boldsymbol{R}_e + \hat{R}_i) - \Psi(\boldsymbol{R}_e - \hat{R}_i) \, .
\end{align}
This Hamiltonian describes two flavors of free, massless Dirac fermions in the low-energy regime, 
since the second term in Eq.~\eqref{Hamiltonian-matrix-2-flavor} vanishes in the continuum limit.

\section{Symmetries in staggered fermions}
\label{sec:symmetries}

The staggered Hamiltonian~\eqref{eq:Hamiltonian-staggerd} possesses a rich symmetry structure, 
reflecting the fact that the spin and flavor degrees of freedom are spatially staggered in this formulation. 
In this section, we clarify the action of discrete spacetime symmetries and internal symmetries, 
both on the staggered fermions and on the Dirac fermions~\eqref{eq:2-2-matrix-fermion} in this system.
We note that staggered fermions in Lagrangian formalism have different degrees of freedom and symmetries.
Their properties are reviewed in Appendix~\ref{sec:appendix_staggered_wilson} for comparison.

\subsection{Discrete symmetries}
\label{subsec:discrete-symmetry}

The lattice Hamiltonian~\eqref{eq:Hamiltonian-staggerd} possesses the following discrete symmetries:
parity symmetry $\mathbb{Z}_2^{\mathrm{P}}$ generated by $\mathrm{P}$,
reflection symmetry $\mathbb{Z}_2^{\mathrm{R}_i}$ generated by $\mathrm{R}_i$ in each spatial direction $(i=x,y,z)$,
time-reversal symmetry $\mathbb{Z}_2^{\mathrm{T}}$ generated by $\mathrm{T}$,
charge-conjugation symmetry $\mathbb{Z}_2^{\mathrm{C}}$ generated by $\mathrm{C}$,
and shift symmetries $\mathbb{Z}_{L_i}^{S}$ in each spatial direction $(i=x,y,z)$ generated by $S_i$,
where $L_i$ denotes the number of lattice sites along the $i$-th direction~\cite{Golterman:2024xos,Li:2024dpq,Chatterjee:2024gje,Aoki:2025vtp}.
These symmetries act on the staggered fermions as
\begin{align}
    \label{eq:lattice-parity-action}
    \mathrm{P}&: \quad \mathrm{P} \chi({\boldsymbol{r}}) \mathrm{P}^{-1}
    = \epsilon(\boldsymbol{r})\, \chi({-\boldsymbol{r}}) \, , 
    \qquad \text{with} \quad \mathrm{P} = \mathrm{R}_x\mathrm{R}_y\mathrm{R}_z \, ,   \\
    \label{eq:lattice-reflection-action}
    \mathrm{R}_i&: \quad \mathrm{R}_i \chi({\boldsymbol{r}}) \mathrm{R}_i^{-1}
    = (-1)^{r_i}\, \chi({\tilde{R}_i(\boldsymbol{r})} )
    = \epsilon(\boldsymbol{r}) \eta_i (\boldsymbol{r})\zeta_i (\boldsymbol{r}) \chi({\tilde{R}_i(\boldsymbol{r})} ) \, , \\
    \label{eq:lattice-time-action}
    \mathrm{T}&: \quad \mathrm{T} \chi({\boldsymbol{r}} )\mathrm{T}^{-1}
    = \epsilon(\boldsymbol{r})\, \chi({\boldsymbol{r}}) \, , 
    \qquad \mathrm{T}\,\i\,\mathrm{T}^{-1} = -\i \, , \\
    \label{eq:lattice-charge-action}
    \mathrm{C}&: \quad \mathrm{C} \chi({\boldsymbol{r}}) \mathrm{C}^{-1}
    = \chi^{\dagger}({\boldsymbol{r}}) \, , \\
    \label{eq:lattice-shift-action}
    {S}_i&: \quad {S}_i \chi({\boldsymbol{r}}) {S}_i^{-1}
    = \zeta_i(\boldsymbol{r})\, \chi({\boldsymbol{r}+\hat{i}}) \, , 
    \qquad {S}_i = T_{\hat{i}}^{(a)} T_{\hat{i}}^{(b)} \, ,
\end{align}
where ${\tilde{R}_i}(r_1, \dots , r_i, \dots , r_3) = (r_1, \dots , -r_i, \dots , r_3)$ and $\epsilon(\boldsymbol{r}) = (-1)^{x + y + z}$.
Here, the spatial coordinate vector of one-component staggered fermions is denoted by $\boldsymbol{r}=(r_1, r_2, r_3) = (x,y,z)$.
The shift operators $S_i$ are composed of $T_{\hat{i}}^{(a)}$ and $T_{\hat{i}}^{(b)}$, which are translation operators acting only on the Majorana fermions $a$ and $b$, respectively:
\begin{align}
\label{eq:translation-op-for-b}
    T_{\hat{i}}^{(a)} a_{\boldsymbol{r}} \left(T_{\hat{i}}^{(a)}\right)^{-1} &= \zeta_i(\boldsymbol{r}) \, a_{\boldsymbol{r} + \hat{i}}\, , 
    \qquad 
    T_{\hat{i}}^{(a)} b_{\boldsymbol{r}} \left(T_{\hat{i}}^{(a)}\right)^{-1} = \zeta_i(\boldsymbol{r}) \, b_{\boldsymbol{r}} \, ,  \notag \\
    T_{\hat{i}}^{(b)} a_{\boldsymbol{r}} \left(T_{\hat{i}}^{(b)}\right)^{-1} &= a_{\boldsymbol{r}} \, , 
    \qquad 
    T_{\hat{i}}^{(b)} b_{\boldsymbol{r}} \left(T_{\hat{i}}^{(b)}\right)^{-1} = \zeta_i(\boldsymbol{r}) \, b_{\boldsymbol{r} + \hat{i}} \, ,
\end{align}
where $\zeta_i(\boldsymbol{r}) := (-1)^{\sum_{k=i+1}^3 r_k}$~\cite{Catterall:2025vrx,Onogi:2025xir}. 

The corresponding actions of these discrete symmetries on the two-flavor Dirac fermion~\eqref{eq:2-2-matrix-fermion} in the infrared continuum theory are given by
\begin{align}
    \label{eq:cont-parity-action}
    \mathrm{P} &: \quad 
    \Psi(\boldsymbol{R}_e) \to (\beta \otimes \mathbbm{1})\, \Psi(-\boldsymbol{R}_e) \, , 
    \qquad \text{with} \quad \mathrm{P} = \mathrm{R}_x\mathrm{R}_y\mathrm{R}_z \, ,  \\
    \label{eq:cont-reflection-X-action}
    \mathrm{R}_x&: \quad 
    \Psi(\boldsymbol{R}_e) \to ( -\beta \gamma_5\alpha_1 \otimes \sigma_1^T)\, \Psi(\tilde{R}_x(\boldsymbol{R}_e))
    = (-\mathrm{i} \beta \alpha_2\alpha_3 \otimes \sigma_1^T)\, \Psi(\tilde{R}_x(\boldsymbol{R}_e)) \, , \\
    \label{eq:cont-reflection-Y-action}
    \mathrm{R}_y&: \quad 
    \Psi(\boldsymbol{R}_e) \to ( \beta \gamma_5  \alpha_2 \otimes  \sigma_2^T  )\, \Psi(\tilde{R}_y(\boldsymbol{R}_e))
    = (\mathrm{i} \beta \alpha_3\alpha_1 \otimes \sigma_2^T)\, \Psi(\tilde{R}_y(\boldsymbol{R}_e))  \, , \\
    \label{eq:cont-reflection-Z-action}
    \mathrm{R}_z&: \quad 
    \Psi(\boldsymbol{R}_e) \to ( -\beta \gamma_5  \alpha_3 \otimes \sigma_3^T )\, \Psi(\tilde{R}_z(\boldsymbol{R}_e))
    = ( -\mathrm{i} \beta \alpha_1\alpha_2 \otimes \sigma_3^T)\, \Psi(\tilde{R}_z(\boldsymbol{R}_e)) \, , \\
    \mathrm{T}&: \quad 
    \Psi(\boldsymbol{R}_e) \to (-\alpha_1 \alpha_3 \otimes \sigma_2^T)\, \Psi(\boldsymbol{R}_e) \, , \\
    \label{eq:cont-lattice-charge-action}
    \mathrm{C}&: \quad 
    \Psi(\boldsymbol{R}_e) \to -(\beta \alpha_3 \alpha_1 \otimes \mathrm{i}\sigma_2^T)\, \Psi^{*}(\boldsymbol{R}_e) \, , \\
    \label{eq:cont-time-action}
    \mathcal{T}&: \quad 
    \Psi(\boldsymbol{R}_e) \to (\alpha_1 \alpha_3 \otimes \mathbbm{1})\, \Psi(\boldsymbol{R}_e) \, , \\
    \label{eq:cont-charge-action}
    \mathcal{C}&: \quad 
    \Psi(\boldsymbol{R}_e) \to \mathrm{i}(\beta \alpha_2 \otimes \mathbbm{1})\, \Psi^{*}(\boldsymbol{R}_e) \, , \\
    \label{eq:cont-shift-action}
    \mathcal{S}_i&: \quad 
    \Psi(\boldsymbol{R}_e) \to (\gamma_5 \otimes \sigma_i^{T})\, \Psi(\boldsymbol{R}_e) \, ,
    \qquad \text{with rotation angle } \frac{\pi}{2} \, .
\end{align}

Here, the symmetry operators $\mathcal{T}$ and $\mathcal{C}$ are defined by
\begin{align}
    \mathcal{T} := -S_z S_x  \mathrm{T} &: \quad 
    \mathcal{T} \chi({\boldsymbol{r}} )\mathcal{T}^{-1}
    = - (-1)^x \chi({\boldsymbol{r}+\hat{x}+\hat{z}}) \, , 
    \qquad \mathcal{T}\,\i\,\mathcal{T}^{-1} = -\i \, , \\
    \mathcal{C} := S_y \mathrm{C} &: \quad 
    \mathcal{C} \chi({\boldsymbol{r}} )\mathcal{C}^{-1}
    = (-1)^z \chi^{\dagger}({\boldsymbol{r}+\hat{y}}) \, .
\end{align}

Note that the last three transformations~\eqref{eq:cont-time-action}--\eqref{eq:cont-shift-action}, generated by $\mathcal{T}$, $\mathcal{C}$, and $\mathcal{S}_i$, are realized only in the continuum limit, 
whereas the parity $\mathrm{P}$ and reflection transformations $\mathrm{R}_i$~\eqref{eq:cont-parity-action}--\eqref{eq:cont-reflection-Z-action} are exact symmetries both on the lattice and in the continuum theory.
There exist two types of time-reversal symmetries, $\mathrm{T}$ and $\mathcal{T}$, as well as two types of charge-conjugation symmetries, $\mathrm{C}$ and $\mathcal{C}$.
Although $\mathcal{T}$ and $\mathcal{C}$ are more natural from the continuum viewpoint, since the definitions of $\mathrm{T}$ and $\mathrm{C}$ involve additional flavor transformations, the latter symmetries play an important role in the discussion of Sec.~\ref{sec:mass-splitting}.
The time-reversal symmetry generated by $\mathcal{T}$ forms a $\mathbb{Z}_4^{\mathcal{T}}$ group, whereas that generated by $\mathrm{T}$ forms a $\mathbb{Z}_2^{\mathrm{T}}$ group.
In particular, the fermion parity $(-1)^F$ is generated by $(\mathcal{T})^2$, which constitutes a subgroup of the vector $\mathrm{U}(1)_V$ symmetry.

\subsection{Conserved charges related to Onsager algebras}
\label{subsec:Onsager-ConservedCharge}

The Hamiltonian~\eqref{eq:Hamiltonian-staggerd} possesses the following conserved charges:
\begin{align}
    \label{eq:vector-charge}
    Q_0 := \sum_{\boldsymbol{r} \in \Lambda} \left(\chi^{\dagger}({\boldsymbol{r}}) \chi({\boldsymbol{r}}) - \frac{1}{2} \right) 
    = \sum_{\boldsymbol{r} \in \Lambda} \frac{\i}{2} a_{\boldsymbol{r}} b_{\boldsymbol{r}} \, ,
\end{align}
which generates the vector $\mathrm{U}(1)_V$ symmetry. In addition, one can define
\begin{align}
    \label{eq:Q-chi-operator}
    Q_{\boldsymbol{\chi}} 
    &:= T_{\boldsymbol{\chi}}^{(b)} Q_0 \left(T_{\boldsymbol{\chi}}^{(b)}\right)^{-1} \notag \\
    &= \frac{1}{2} \sum_{\boldsymbol{r} \in \Lambda} 
    \left(\zeta_y(\boldsymbol{r})\right)^{n_y} 
    \left(\zeta_x(\boldsymbol{r})\right)^{n_x}
    \left(\chi({\boldsymbol{r}}) + \chi^{\dagger}({\boldsymbol{r}}) \right)
    \left(\chi({\boldsymbol{r}+\boldsymbol{\chi}}) - \chi^{\dagger}({\boldsymbol{r}+\boldsymbol{\chi}}) \right) \notag \\
    &= \frac{\i}{2} \sum_{\boldsymbol{r} \in \Lambda} 
    \left(\zeta_y(\boldsymbol{r})\right)^{n_y} 
    \left(\zeta_x(\boldsymbol{r})\right)^{n_x} 
    a_{\boldsymbol{r}} b_{\boldsymbol{r} + \boldsymbol{\chi}} \, .
\end{align}
Here, $T_{\boldsymbol{\chi}}^{(b)}$ denotes a translation operator analogous to Eq.~\eqref{eq:translation-op-for-b}, whose action is extended to an arbitrary lattice vector $\boldsymbol{\chi} = n_x \hat{x} + n_y \hat{y} + n_z \hat{z}$ with $n_i \in \mathbb{Z}$,
\begin{align}
    T_{\boldsymbol{\chi}}^{(b)} a_{\boldsymbol{r}} \left(T_{\boldsymbol{\chi}}^{(b)}\right)^{-1} &= a_{\boldsymbol{r}} \ , \notag \\
    T_{\boldsymbol{\chi}}^{(b)} b_{\boldsymbol{r}} \left(T_{\boldsymbol{\chi}}^{(b)}\right)^{-1} &=
    \left(\zeta_y(\boldsymbol{r})\right)^{n_y} 
    \left(\zeta_x(\boldsymbol{r})\right)^{n_x} 
    b_{\boldsymbol{r} + \boldsymbol{\chi}} \ .
\end{align}
Importantly, the conserved charges $Q_0$ and $Q_{\boldsymbol{\chi}}$ generate the Onsager algebra ($\mathrm{Ons}3$)~\cite{Aoki:2025vtp}.

Remarkably, in the continuum limit, the conserved charges $Q_{{i}}$ associated with a single lattice shift $\hat{i} = \hat{x}, \hat{y}, \hat{z}$,
\begin{align}
    \label{eq:Qi-ConservedCharges-satisfying-OnsagerAlgebras}
    Q_{{i}} 
    = T_{\hat{i}}^{(b)} Q_0 \left(T_{{\hat{i}}}^{(b)}\right)^{-1} \, ,
\end{align}
act on the two-flavor Dirac fermion~\eqref{eq:2-2-matrix-fermion} as
\begin{align}
    \label{eq:SU2A-transform}
    \lim_{L \to \infty}[Q_{{i}}, \Psi(\boldsymbol{R}_e)] 
    = (\gamma_5 \otimes \sigma_i^T) \, \Psi(\boldsymbol{R}_e) \, .
\end{align}
Here, $\gamma_5$ acts on the spin indices, while $\sigma_i$ acts on the flavor indices of the fermion.

These conserved charges correspond to the broken generators of $\mathrm{SU}(2)_L \times \mathrm{SU}(2)_R$, 
acting with opposite signs on the left- and right-handed Weyl fermions.
Each conserved charge $Q_{\hat{i}}$ may thus be regarded as the generator of a $\mathrm{U}(1)_{F_i}$ subgroup of 
$\mathrm{SU}(2)_L \times \mathrm{SU}(2)_R \times \mathrm{U}(1)_A$~\cite{Onogi:2025xir}.

\section{Classification of mass terms for the staggered fermion }
\label{sec:classification-mass}

As we have seen in Sec.~\ref{sec:Hamiltonian-lattice}, in the staggered fermion Hamiltonian~\eqref{eq:Hamiltonian-staggerd}, a Dirac fermion is reconstructed from eight one-component staggered fermions residing within a unit cube.
Accordingly, the Dirac mass terms are not restricted to on-site bilinears defined at the same lattice site inside the unit cube; they may also arise from hopping terms that extend over one-link, two-link, or three-link separations within the cube.
In this section, we classify all possible bilinear Dirac mass terms that are Hermitian, particle-number-preserving, local within a unit cube, and gap out the continuum Dirac fermions.
The symmetry properties of these mass terms are summarized in Table~\ref{table:mass-symmetry}.
As we mentioned, it should be noted that the staggered fermions in Lagrangian formalism give a different system from that in Hamiltonian formalism.
The case of Lagrangian formalism will be briefly summarized in Appendix~\ref{sec:appendix_staggered_wilson}.

\begin{table}[htbp]
    \begin{center}
        \scalebox{1.1}{
    \begin{tabular}{|c|c|c|c|c|c|c|c|c|c|c|c|c|c|c|c|}\hline
                       & $\mathrm{P}$ & $\mathrm{R}_x$ & $\mathrm{R}_y$ & $ \mathrm{R}_z$ & $\mathrm{T}$ & $\mathrm{C}$ & $\mathcal{T}$ & $\mathcal{C}$ & ${S}_x $  & ${S}_y$  & ${S}_z$  & ${Q}_0$  & ${Q}_x$  & ${Q}_y$  & ${Q}_z$   \\ \hline
    $\delta H^{(0)}$   & $\circ$      & $\circ$        & $\circ$        & $\circ$         & $\circ$      & $\times$     & $\circ$       & $\circ$       & $\times$  & $\times$ & $\times$ & $\circ$  & $\times$ & $\times$ & $\times$\\ \hline
    $\delta H^{(1)}_x$ & $\times$     & $\times$       & $\circ$        & $\circ$         & $\circ$      & $\circ$      & $\times$      & $\circ$       & $\times$  & $\circ$  & $\circ$  & $\circ$  & $\times$ & $\circ$  & $\circ$\\ \hline
    $\delta H^{(1)}_y$ & $\times$     & $\circ$        & $\times$       & $\circ$         & $\circ$      & $\circ$      & $\times$      & $\times$      & $\times$  & $\times$ & $\circ$  & $\circ$  & $\times$ & $\times$ & $\circ$\\ \hline
    $\delta H^{(1)}_z$ & $\times$     & $\circ$        & $\circ$        & $\times$        & $\circ$      & $\circ$      & $\times$      & $\times$      & $\times$  & $\times$ & $\times$ & $\circ$  & $\times$ & $\times$ & $\times$\\ \hline
    $\delta H^{(2)}_x$ & $\circ$      & $\circ$        & $\times$       & $\times$        & $\times$     & $\circ$      & $\circ$       & $\times$      & $\times$  & $\times$ & $\times$ & $\circ$  & $\times$ & $\times$ & $\times$\\ \hline
    $\delta H^{(2)}_y$ & $\circ$      & $\times$       & $\circ$        & $\times$        & $\times$     & $\circ$      & $\circ$       & $\times$      & $\times$  & $\times$ & $\times$ & $\circ$  & $\times$ & $\times$ & $\times$\\ \hline
    $\delta H^{(2)}_z$ & $\circ$      & $\times$       & $\times$       & $\circ$         & $\times$     & $\circ$      & $\circ$       & $\times$      & $\times$  & $\times$ & $\times$ & $\circ$  & $\times$ & $\times$ & $\times$\\ \hline
    $\delta H^{(3)}$   & $\times$     & $\times$       & $\times$       & $\circ$         & $\circ$      & $\times$     & $\circ$       & $\times$      & $\times$  & $\times$ & $\times$ & $\circ$  & $\times$ & $\times$ & $\times$\\ \hline
\end{tabular}}
\caption{Commutation relation between masses and symmetries}
\label{table:mass-symmetry}
$\circ$ : commutative , $\times$ : non-commutative.
\end{center}
\end{table}


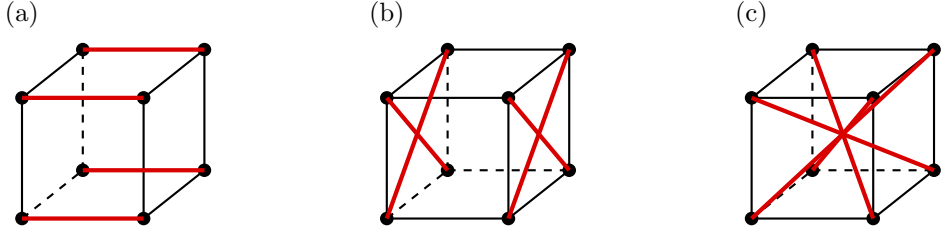
\begin{figure}[htbp]
    \centering
    \begin{tikzpicture}[
        scale=1.6,
        x={(1cm, 0cm)}, y={(0.5cm, 0.4cm)}, z={(0cm, 1cm)},
        vertex/.style={circle, fill=black, inner sep=1.8pt},
        edge/.style={thick, black},
        hidden edge/.style={thick, dashed, black},
        hopping/.style={ultra thick, red!85!black}
    ]

    \def\drawcube{
        \draw[hidden edge] (0,0,0) -- (0,1,0);
        \draw[hidden edge] (0,1,0) -- (1,1,0);
        \draw[hidden edge] (0,1,0) -- (0,1,1);
        
        \draw[edge] (0,0,0) -- (1,0,0) -- (1,0,1) -- (0,0,1) -- cycle; 
        \draw[edge] (1,0,0) -- (1,1,0) -- (1,1,1) -- (1,0,1); 
        \draw[edge] (0,0,1) -- (0,1,1) -- (1,1,1); 
        
        \foreach \x in {0,1} {
            \foreach \y in {0,1} {
                \foreach \z in {0,1} {
                    \node[vertex] at (\x,\y,\z) {};
                }
            }
        }
    }

    \begin{scope}[xshift=0cm]
        \node[anchor=south, font=\small] at (0.0, 0.0, 1.5) {(a)};
        \drawcube
        \draw[hopping] (0,0,0) -- (1,0,0);
        \draw[hopping] (0,1,0) -- (1,1,0);
        \draw[hopping] (0,0,1) -- (1,0,1);
        \draw[hopping] (0,1,1) -- (1,1,1);
    \end{scope}

    \begin{scope}[xshift=3cm]
        \node[anchor=south, font=\small] at (0.0, 0.0, 1.5) {(b)};
        \drawcube
        \draw[hopping] (0,0,1) -- (0,1,0); 
        \draw[hopping] (0,1,1) -- (0,0,0); 
        \draw[hopping] (1,0,1) -- (1,1,0); 
        \draw[hopping] (1,1,1) -- (1,0,0); 
    \end{scope}

    \begin{scope}[xshift=6cm]
        \node[anchor=south, font=\small] at (0.0, 0.0, 1.5  ) {(c)};
        \drawcube
        \draw[hopping] (0,0,0) -- (1,1,1);
        \draw[hopping] (1,0,0) -- (0,1,1);
        \draw[hopping] (0,1,0) -- (1,0,1);
        \draw[hopping] (0,0,1) -- (1,1,0);
    \end{scope}

    \end{tikzpicture}
    \caption{Schematic representation of the hoppings within a unit cube corresponding to the different mass terms: (a) one-link mass $\delta H_x^{(1)}$, (b) two-link mass $\delta H_x^{(2)}$, and (c) three-link mass $\delta H^{(3)}$. The black dots represent the lattice sites, and the thick red lines denote the hopping interactions.}
    \label{fig:mass_terms_hoppings}
\end{figure}

\subsection{On-site mass}
\label{sec:zero-link-mass}

The on-site mass term of staggered fermions yields the Dirac scalar mass,
\begin{align}
    \label{eq:zero-link-mass}
    \delta H^{(0)} &= \sum_{\boldsymbol{r}}  m \epsilon (\boldsymbol{r})   \chi^{\dagger} (\boldsymbol{r}) \chi (\boldsymbol{r} )     \notag \\
    &= \frac{m}{2}\sum_{\boldsymbol{R}_e}  \Psi ^{\dagger} (\boldsymbol{R}_e) ( \gamma_0   \otimes \mathbbm{1}  )  \Psi  (\boldsymbol{R}_e) \, .
\end{align}
As shown in Table~\ref{table:mass-symmetry}, this term explicitly breaks the single-shift symmetries as well as charge conjugation symmetry,
while it manifestly preserves the full three-dimensional rotational (cubic) symmetry of the lattice.

\subsection{One-link mass}
\label{subsec:one-link-mass}

Within the unit cube, as shown in the left panel (a) in Figure.~\ref{fig:mass_terms_hoppings}, one can construct the following three independent one-link mass terms,
\begin{align}
    \label{eq:one-link-mass}
    \delta H^{(1)}_i &= \sum_{\boldsymbol{r}} \mathrm{i} m P_{i+}(\boldsymbol{r}) \left[ \chi^{\dagger} (\boldsymbol{r}) \chi (\boldsymbol{r} + \hat{i}) + \chi (\boldsymbol{r}) \chi ^{\dagger} (\boldsymbol{r} + \hat{i})  \right]  \notag \\
    &= \frac{m}{2}\sum_{\boldsymbol{R}_e}  \Psi ^{\dagger} (\boldsymbol{R}_e) (\mathrm{i} \gamma_0 \gamma_5  \otimes \sigma_i  )  \Psi  (\boldsymbol{R}_e) \, ,
\end{align}
where the projection operators are defined as
\begin{align}
    P_{i\pm} (\boldsymbol{r}) = \frac{1}{2} (1 \pm (-1)^{r_i})  \, .
\end{align}
These terms correspond to pseudo-scalar masses and thus break parity symmetry. 
Furthermore, unlike an on-site mass term, the choice of a specific spatial direction for the one-link hopping explicitly breaks the three-dimensional rotational symmetry (cubic symmetry) of the lattice.
The rotational symmetry acts on the one-component staggered fermions.

However, as summarized in Table~\ref{table:mass-symmetry}, these mass terms preserve more symmetries than other possible Dirac mass terms.

Notably, $\delta H_x^{(1)}$ respects many of the symmetries of the staggered fermion Hamiltonian~\eqref{eq:Hamiltonian-staggerd},
with the exception of $P$, $R_x$, $S_x$, $Q_x$ as well as $\mathcal{T}$.
This invariance implies the absence of a mixed 't~Hooft anomaly between the vector $\mathrm{U}(1)_V$ symmetry generated by $Q_0$ and the non-singlet $\mathrm{U}(1)_{F_i}$ symmetries generated by $Q_y$ and $Q_z$ (see Sec.~\ref{subsec:Onsager-ConservedCharge}). 
We note that this particular conclusion relies on the "normal definition" for the staggered phases~\eqref{eq:staggered-phase}.
If a different definition obtained by
even permutations (the $\mathbb{Z}_3$ cyclic group) of the coordinate labels $\{x,y,z\}$ were adopted, a different one-link mass term would preserve the largest residual symmetry.~\footnote{
\label{footnote:permutation}
One could redefine the phase by permuting the coordinate labels $\{x,y,z\}$ under the symmetric group~$S_3$. Even permutations (the $\mathbb{Z}_3$ cyclic group) represent a trivial relabeling.
Odd permutations are also allowed, but they fail to produce standard finite-difference forms for the kinetic terms. Consequently, our analysis is strictly restricted to the ``normal definition" under the current coordinate convention.}

Furthermore, this symmetry structure suggests that in a $3+1$~D staggered fermion system with a $2+1$~D boundary—characterized by a kink mass profile of the one-link mass $\delta H_x^{(1)}$—the charges $Q_y$ and $Q_z$ remain conserved even within the boundary theory. We will further discuss the physical implications of these conserved charges in the lower-dimensional boundary theory in Sec.~\ref{sec:mass-splitting}.

\subsection{Two-link mass}
\label{subsec:two-link-mass}

The two-link mass terms yield the following Dirac flavor masses,
\begin{align}
    \delta H^{(2)}_x 
    &=  4 \mathrm{i}m \sum_{\boldsymbol{r}}(-1)^{x} P_{y+}(\boldsymbol{r}) \chi^{\dagger} (\boldsymbol{r}) \left[P_{z+}(\boldsymbol{r}) \chi (\boldsymbol{r}+ \hat{y} + \hat{z}) + P_{z-}(\boldsymbol{r}) \chi (\boldsymbol{r}+ \hat{y} - \hat{z})  \right] + \text{h.c.} \notag \\
    &= \sum_{\boldsymbol{R}_e} m \Psi ^{\dagger} (\boldsymbol{R}_e) ( \gamma_0  \otimes \sigma_1  )  \Psi  (\boldsymbol{R}_e)  \, , \\
    \delta H^{(2)}_y 
    &=  4 \mathrm{i}m\sum_{\boldsymbol{r}}  (-1)^y P_{x+}(\boldsymbol{r}) \chi^{\dagger} (\boldsymbol{r})  \left[ P_{z+}(\boldsymbol{r}) \chi (\boldsymbol{r}+ \hat{x} + \hat{z}) + P_{z-} (\boldsymbol{r})\chi (\boldsymbol{r}+ \hat{x} - \hat{z})  \right]  + \text{h.c.} \notag \\
    &= \sum_{\boldsymbol{R}_e} m \Psi ^{\dagger} (\boldsymbol{R}_e) ( \gamma_0  \otimes \sigma_2  )  \Psi  (\boldsymbol{R}_e)  \, , \\
    \delta H^{(2)}_z 
    &= 4 \mathrm{i}m \sum_{\boldsymbol{r}}  (-1)^{z} P_{x+}(\boldsymbol{r})\chi^{\dagger} (\boldsymbol{r}) \left[  P_{y+} (\boldsymbol{r}) \chi (\boldsymbol{r}+ \hat{x} + \hat{y}) + P_{y-}(\boldsymbol{r}) \chi (\boldsymbol{r}+ \hat{x} - \hat{y})  \right] + \text{h.c.} \notag \\
    &= \sum_{\boldsymbol{R}_e} m \Psi ^{\dagger} (\boldsymbol{R}_e) ( \gamma_0  \otimes \sigma_3  )  \Psi  (\boldsymbol{R}_e)  \,  .
\end{align}
As shown in the center panel in Fig.~\ref{fig:mass_terms_hoppings}, each mass term $\delta H_i^{(2)}$ is constructed by diagonal hoppings on the plane perpendicular to the $i$-direction within the unit cube. 
Just as in the one-link case, selecting a specific plane for these hoppings explicitly breaks the three-dimensional rotational (cubic) symmetry.

In contrast to the one-link masses discussed in Sec.~\ref{subsec:one-link-mass}, which are pseudo-scalar masses and thus break parity symmetry, the two-link masses are Dirac-type flavor masses and therefore preserve parity symmetry as summarized in Table~\ref{table:mass-symmetry}. Furthermore, while the time-reversal symmetry generated by $\mathrm{T}$ and the shift symmetries generated by $S_x$ and $S_z$ of the staggered fermions are explicitly broken, a modified time-reversal symmetry generated by $\mathcal{T}$---which is a combination of $\mathrm{T}$, $S_x$, and $S_z$---remains preserved.

\subsection{Three-link mass}
\label{subsec:three-link-mass}

The three-link term gives rise to a Dirac pseudo-scalar mass,
\begin{align}
    \delta H^{(3)} 
    &= 4m\sum_{\boldsymbol{r}} \left[ -P_{x+}(\boldsymbol{r}) P_{y+}(\boldsymbol{r}) \chi ^{\dagger} (\boldsymbol{r}) \left( P_{z+}(\boldsymbol{r}) \chi (\boldsymbol{r} +\hat{x} + \hat{y} + \hat{z}  ) - P_{z-}(\boldsymbol{r}) \chi (\boldsymbol{r} +\hat{x} + \hat{y} - \hat{z}  )  \right) \right. \notag \\
    & \quad \left.  +P_{x-}(\boldsymbol{r}) P_{y+}(\boldsymbol{r}) \chi ^{\dagger} (\boldsymbol{r}) \left( P_{z+}(\boldsymbol{r}) \chi (\boldsymbol{r} -\hat{x} + \hat{y} + \hat{z}  ) - P_{z-}(\boldsymbol{r}) \chi (\boldsymbol{r} -\hat{x} + \hat{y} - \hat{z}  )  \right)    \right] + \text{h.c.} \notag \\
    &= \sum_{\boldsymbol{R}_e} m \Psi ^{\dagger} (\boldsymbol{R}_e) ( \mathrm{i} \gamma_0 \gamma_5  \otimes \mathbbm{1}  )  \Psi  (\boldsymbol{R}_e)  \, .
\end{align}
This mass term is constructed by diagonal hoppings connecting all sites across the main diagonals of the unit cube as shown in the right panel in Figure.~\ref{fig:mass_terms_hoppings}.
Although it couples sites symmetrically across the cube, it only preserves specific spatial reflection such as $\mathrm{R}_z$. Because other spatial reflections are broken, the full three-dimensional rotational (cubic) symmetry is explicitly broken.
Consequently, as summarized in Table~\ref{table:mass-symmetry}, this term breaks almost all the symmetries of the staggered Hamiltonian~\eqref{eq:Hamiltonian-staggerd}.

\section{Dimensional reduction by mass splitting}
\label{sec:mass-splitting}


In Sec.~\ref{sec:classification-mass}, we classified all possible Dirac fermion mass terms in the staggered fermion Hamiltonian
that are Hermitian, particle-number-preserving and local
within a unit cell.
Among them, the one-link mass $\delta H_x^{(1)}$ in the $x$-direction~\eqref{eq:one-link-mass} is distinguished by its exceptionally large symmetry compared to the other mass terms.
In particular, the mass term commutes with not only $Q_0$ but also $Q_y$ and $Q_z$ given by Eq.~\eqref{eq:Qi-ConservedCharges-satisfying-OnsagerAlgebras}.
In this section, we focus on this term, $\delta H_x^{(1)}$, and introduce a kink mass with the profile given in Eq.~\eqref{eq:mass-profile}. 
We then derive the edge modes localized on the resulting domain wall.

Finally, we discuss the physical interpretation of the conserved charges $Q_y$ and $Q_z$,
which satisfy the Onsager algebra, and analyze the mixed 't~Hooft anomalies between spacetime reflection symmetries and the symmetries generated by $Q_i$ on the domain wall.

\subsection{Derivation of Domain-wall fermions}
\label{subsec:Derivation-DW-fermions}
We assume the lattice size for $x$-direction is given by 
$L_x = 2N_x$ with $N_x$ even, so that $L_x \in 4 {\mathbb Z}$,
to impose the periodic boundary condition on the following system.
We start with a Hamiltonian given by
\begin{align}
    H_{DW} &= H + M_x^{(1)} \, ,
\end{align}
where
$H$ is defined by Eq.~\eqref{eq:Hamiltonian-staggerd}, and the  mass term $M_x$ is
\begin{align}
    \label{eq:Kink-mass}
   M_x^{(1)} &= \i  \sum_{\boldsymbol{R}} \frac{m_0}{2} \theta(R_x) \Psi ^{\dagger} (\boldsymbol{R}_e) ( \gamma_0 \gamma_5  \otimes \sigma_1  )  \Psi  (\boldsymbol{R}_e) \, ,
\end{align}
with
\begin{align}
    \label{eq:mass-profile}
    \theta(R_x) &= \begin{cases}
        1 & 0 \leq R_x \leq \frac{N_x}{2}-1  \, , \\
        -1 & \frac{N_x}{2} \leq R_x \leq N_x  \, . 
    \end{cases}
\end{align}
The full Hamiltonian can be expressed as
\begin{align}
    \label{eq:DW-Hamiltonian}
    H_{DW} &= H + M_x^{(1)} \notag \\
    &= \i \sum_{\boldsymbol{r}} \Psi^\dagger(\boldsymbol{r}) \left[
        (\alpha_i \otimes \mathbbm{1}_{2 \times 2}) \frac{\nabla_i}{2}  
        + (\beta \gamma_5 \otimes \sigma_i^T) \frac{\nabla_i^2}{2} \right] \Psi(\boldsymbol{r}) \notag \\
        &  \quad + \i  \sum_{\boldsymbol{r}}
        \frac{m_0}{2}\theta(x)\Psi ^{\dagger} (\boldsymbol{r}) ( \beta \gamma_5  \otimes \sigma_1  )  \Psi  (\boldsymbol{r}) \, .
\end{align}
where for simplicity, we have denoted the spatial coordinate 
$\boldsymbol{R}_e$ by $\boldsymbol{r}$ and $R_x$ by $x$.
Since the Hamiltonian~\eqref{eq:DW-Hamiltonian} has two domain-walls at $x=0$ and $x=\frac{N_x}{2}$,
we can expect that, in this system, zero modes are localized around the walls in the low-energy limit \cite{Kaplan:1992bt,Shamir:1993zy}.
Let us derive the zero modes in what follows.

To this end,
we now perform a Wick rotation and work in Euclidean spacetime.
The Hamiltonian density is defined by
\begin{align}
    \label{eq:general-Hamiltonian-Euclidean}
    H &= \sum_{\boldsymbol{r}} \Psi^\dagger(\boldsymbol{r}) \hat{h}_E \Psi(\boldsymbol{r}) \, ,
\end{align}
where
\begin{align}
    \label{eq:hamiltonian-density}
    \hat{h}_E &= \left[
        (\alpha_i \otimes \mathbbm{1}_{2 \times 2}) \frac{\nabla_i}{2}  
        + \mathrm{i}\beta \Gamma_M^i \frac{\nabla_i^2}{2} + \mathrm{i} \beta \Gamma_M^1 \frac{m_0 \theta(x)}{2} \right] \, . 
\end{align}
The Euclidean gamma matrices $\gamma_\mu^E$ are defined by
\begin{align}
    \gamma_0^M = \gamma_0^E \, , \, \i  \gamma_{i}^M = \gamma_{i}^E \, , \, \gamma_5 ^M = \gamma_5^E \, , \, \i\alpha_i ^{ M} = \alpha_i^{E} \, , 
\end{align}
so that they satisfy $\{ \gamma_\mu ^E , \gamma_\nu ^E \} = 2 \delta_{\mu\nu}$,
and we have omitted the  
superscript $E$ for simplicity.
$\Gamma_M^i$ is a tensor matrix that is defined as
\begin{align}
    \label{eq:Def-GammaM}
    \Gamma_M^i & := 
         \gamma_5 \otimes \sigma_i ^T    \, .
\end{align}
Then, the action associated with the Hamiltonian~\eqref{eq:general-Hamiltonian-Euclidean} is given by\footnote{
In Minkowski spacetime, the action associated to the Hamiltonian~\eqref{eq:general-Hamiltonian-Euclidean} is given by
\begin{align}
    S_M = \int dt \sum_{\boldsymbol{r}} ( \Psi ^\dagger \i \partial_0 \Psi -  \Psi ^\dagger \hat{h}_E \Psi ) \ ,
\end{align}
that is transformed under the Wick rotation as follows,
\begin{align}
    S_M = -\i \int dt \sum_{\boldsymbol{r}} ( -\Psi ^\dagger \partial_0 \Psi -  \Psi ^\dagger \hat{h}_E \Psi ) \ .
\end{align}
Hence, the action in Euclidean spacetime~\eqref{eq:action--matrix-2-flavor-Euclidian} can be defined as
\begin{align}
    S := -iS_M = \int dt \sum_{\boldsymbol{r}} ( \Psi ^\dagger \partial_0 \Psi +  \Psi ^\dagger \hat{h}_E \Psi ) \, .
\end{align}
}
\begin{align}
\label{eq:action--matrix-2-flavor-Euclidian}
    S &=  \int dt \sum_{\boldsymbol{r}}  {\overline{\Psi}}(t,\boldsymbol{r}) \left(  \gamma^0 \partial_0 + \gamma^0 \hat{h}_E \right) {\Psi} (t,\boldsymbol{r}) \notag \\
    &= \int dt \sum_{\boldsymbol{r}}  {\overline{\Psi}}(t,\boldsymbol{r}) \left( \gamma^0 \partial_0 + \gamma^i \frac{\nabla_i}{2} +  \mathrm{i} \Gamma_M^i \frac{\nabla_i^2}{2} + \mathrm{i} \Gamma_M^1 \frac{m_0 \theta(x)}{2}    \right)   {\Psi} (t,\boldsymbol{r})    \, . 
\end{align}
with ${\overline{\Psi}} := {\Psi}^\dagger \gamma^0$.
After performing a Fourier transformation in the $t$, $y$, and $z$ directions,
the action becomes
\begin{align}
    S&=   \int dp_0 \sum_{(p_y,p_z)\in \text{BZ}} {\overline{\Psi}}(p,x) \left[ \slashed{\mathcal{D}}^{tyz}(p,x) + \slashed{D}^x (p,x)  \right] {\Psi} (p,x) \, ,
\end{align}
where the momentum $p$ denotes $(p_0, p_y, p_z)$ and $\text{BZ}$ is the Brillouin zone in $y$ and $z$ directions defined as
$0 \le k_y, k_z < \pi$.
Here each Dirac operator is given by
\begin{align}
    \label{eq:DiracOP-tyz}
    \slashed{\mathcal{D}}^{tyz}(p) &= \i \slashed{p}_0 + \sum_{i=2,3}\left( \i \gamma_i \sin (p_i) - \mathrm{i} \Gamma_M^i (1- \cos (p_i)) \right) \notag \\
    &= \i \slashed{p}_0 + \sum_{i=2,3}\left( \i \gamma_i \sin (p_i) -  \i (\gamma_5 \otimes \sigma_i) ^T (1- \cos (p_i)) \right)  \, , \\
    \label{eq:DiracOP-x}
    \slashed{D}^x (p,x) &= \gamma_1 \frac{\nabla_1}{2} + \Gamma_M^1 \left( \frac{\nabla_1^2}{2} + \frac{m_0 \theta(x)}{2}  \right) \notag \\
    &= \left\{ - (\sigma_1 \otimes \sigma_2) \otimes \mathbbm{1}  \right\} \frac{\nabla_1}{2} + \mathrm{i}\left\{  (\mathbbm{1} \otimes \sigma_3) \otimes \sigma_1  \right\} \left( \frac{\nabla_1^2}{2} + \frac{m_0 \theta(x)}{2}  \right) \, .
\end{align}

The equation of motion for the fermion is given by
\begin{align}
    \left[ \slashed{\mathcal{D}}^{tyz}(p,x) + \slashed{D}^x (p,x)  \right] {\Psi} (p,x) &=0 \, . 
\label{eq:DW_EOM}
\end{align}
The solution of Eq.~\eqref{eq:DW_EOM} is also the zero mode of the positive semi-definite operator
\begin{eqnarray}
    \left[ \slashed{\mathcal{D}}^{tyz}(p,x) + \slashed{D}^x (p,x)  \right]^{\dagger} \left[ \slashed{\mathcal{D}}^{tyz}(p,x) + \slashed{D}^x (p,x)  \right] .
\end{eqnarray}
Due to the anti-commutativity of the gamma matrices, it is easy to find that this operator can be decomposed into the sum of positive semi-definite operators as
\begin{align}
    &
    \left[ \slashed{\mathcal{D}}^{tyz}(p,x) + \slashed{D}^x (p,x)  \right]^{\dagger} \left[ \slashed{\mathcal{D}}^{tyz}(p,x) + \slashed{D}^x (p,x)  \right] 
    \notag \\ 
   &
    = (\slashed{\mathcal{D}}^{tyz}(p,x)) ^{\dagger} \slashed{\mathcal{D}}^{tyz}(p,x) + (\slashed{D}^x (p,x))^{\dagger} \slashed{D}^x (p,x),
    \notag \\
    &= \left( p_0 ^2 + \sum_{i=2,3} ( \sin^2 (p_i) + (1- \cos(p_i))^2 )  \right) - \left( \left( \frac{\nabla_1}{2} \right)^2 -  (u(x)) ^2 + \mathrm{i}\gamma_1 \Gamma_M^1 \left( \frac{\nabla_1 u}{2} \right)  \right) \, ,
\end{align}
where
\begin{align}
    \label{eq:modified-mass}
    u(x) = \frac{\nabla_1^2}{2} + \frac{m_0 \theta(x)}{2} \, .
\end{align}
The solution of Eq.~\eqref{eq:DW_EOM} is a zero mode of the first and second terms simultaneously.
Therefore, in order to extract the localized zero modes 
on the domain-wall, we consider the reduced equation
\begin{align}
    \label{eq:eomX-of-Psi}
  {\slashed{D}^x}  (p,x)  {\Psi} (p,x) &= 0 \, .
\end{align} 

Introducing the unitary matrix
\begin{align}
    \label{eq:unitary-matrix-Lambda}
    \tilde{\Lambda} &:= (\Lambda_1\otimes \mathbbm{1} ) \otimes  \Lambda_1 \, ,
\end{align}
where
\begin{align}
    \Lambda_1 &= \Lambda_1^{-1} = \frac{1}{\sqrt{2}} (\sigma_1 + \sigma_3) =  \frac{1}{\sqrt{2}} \begin{pmatrix}
        1 & 1 \\
        1 & -1
    \end{pmatrix} \, ,  \\
    &\Lambda_1 \sigma_1 \Lambda_1^{-1} = \sigma_3 \, , \quad   \Lambda_1 \sigma_2 \Lambda_1^{-1} = -\sigma_2 \, , \quad \Lambda_1 \sigma_3 \Lambda_1^{-1} = \sigma_1 \, ,  
\end{align}
the Dirac operator associated with the $x$-direction in Eq.~\eqref{eq:eomX-of-Psi} can be diagonalized as
\begin{align}
    \label{eq:diagonalized-DiracOP}
    \widetilde{\slashed{D}^x} (p,x) \widetilde{\Psi} (p,x)&= \left[ \left\{ - (\sigma_3 \otimes \sigma_2) \otimes \mathbbm{1}  \right\} \frac{\nabla_1}{2} + \mathrm{i}\left\{  (\mathbbm{1} \otimes \sigma_3) \otimes \sigma_3  \right\} u(x)\right] \widetilde{\Psi} (p,x) =0 \, ,
\end{align}
where
\begin{align}
    \label{eq:tilde-Psi}
    \widetilde{\Psi} (p,x) = \tilde{\Lambda}{\Psi} (p,x) \, .
\end{align}
By multiplying $\Gamma_2$, defined in Eq.~\eqref{eq:def:Gamma12}, from the left on Eq.~\eqref{eq:diagonalized-DiracOP}, the equation can be rewritten in the simplified form
\begin{align}
     \label{eq:rewritten-eomX-of-tilde-Psi}
     P_{-} \widetilde{\Psi} (p,x+1) +  P_{+} \widetilde{\Psi} (p,x-1) + m'  \widetilde{\Psi} (p,x) = 0 \, ,
\end{align}
where
\begin{align}
    \label{eq:def:Gamma12}
    \Gamma_1 &= - (\sigma_3 \otimes \sigma_2) \otimes \mathbbm{1} \, , \,  \Gamma_2 = (\mathbbm{1} \otimes \sigma_3) \otimes \sigma_3  \, , \\
    \mathcal{P}_{a/b} &=\frac{1}{2} \begin{pmatrix}
            1 & \mp \sigma_3 \\
            \mp \sigma_3 &  1 
        \end{pmatrix} \, , \\
    P_{\pm} & = \frac{1}{2} \left( 1 \pm \mathrm{i} \Gamma_2 \Gamma_1    \right) 
    = \begin{pmatrix}
        \mathcal{P}_{a/b} & 0 \\
        0 &  \mathcal{P}_{b/a}
    \end{pmatrix}  \,  ,
\end{align}
and 
\begin{align}
    m'(x) &=  a m_0 \theta (x) -2 \, ,
\end{align}
with the lattice spacing $a$.
From Eq.~\eqref{eq:rewritten-eomX-of-tilde-Psi}, we obtain
\begin{align}
    \label{eq:eom-for-mathcal-Pb-plus}
    \mathcal{P}_b \left[ \widetilde{\Psi}^{(1)} (p,x+1) + m'  \widetilde{\Psi}^{(1)} (p,x) \right] & = 0 \, , \\
    \label{eq:eom-for-mathcal-Pa-plus}
    \mathcal{P}_a \left[ \widetilde{\Psi}^{(2)} (p,x+1) + m'  \widetilde{\Psi}^{(2)} (p,x) \right] & = 0 \, , \\
    \label{eq:eom-for-mathcal-Pa-minus}
    \mathcal{P}_a \left[ \widetilde{\Psi}^{(1)} (p,x-1) + m'  \widetilde{\Psi}^{(1)} (p,x) \right] & = 0 \, , \\
    \label{eq:eom-for-mathcal-Pb-minus}
    \mathcal{P}_b \left[ \widetilde{\Psi}^{(2)} (p,x-1) + m'  \widetilde{\Psi}^{(2)} (p,x) \right] & = 0 \, .
\end{align}

The following unitary matrices
\begin{align}
    \mathcal{U}_a = \frac{1}{\sqrt{2}} \begin{pmatrix}
        1 & \sigma_3 \\
        1 & -\sigma_3
    \end{pmatrix} \, , \, \mathcal{U}_a^{-1} = \frac{1}{\sqrt{2}} \begin{pmatrix}
        1 & 1 \\
        \sigma_3 & -\sigma_3
    \end{pmatrix} \, , \\
    \mathcal{U}_b = \frac{1}{\sqrt{2}} \begin{pmatrix}
        1 & -\sigma_3 \\
        1 & \sigma_3
    \end{pmatrix} \, , \,\mathcal{U}_b^{-1} = \frac{1}{\sqrt{2}} \begin{pmatrix}
        1 & 1\\
        -\sigma_3  & \sigma_3
    \end{pmatrix} \, ,
\end{align}
diagonalize the matrices $\mathcal{P}_{a/b}$ as
\begin{align}
    \mathcal{U}_{a/b} \mathcal{P}_{a/b} \mathcal{U}_{a/b}^{-1} & = \begin{pmatrix}
         0 & 0 \\
         0 & 1 
    \end{pmatrix} \, .
\end{align}
In this new basis, the fermion in Eq.~\eqref{eq:tilde-Psi} is transformed as
\begin{align}
    \label{eq:Psi-transformed-by-U}
    \begin{pmatrix}
        \mathcal{U}_a \widetilde{\Psi}^{(1)} \\
        \mathcal{U}_b \widetilde{\Psi}^{(2)}
    \end{pmatrix} & = \begin{pmatrix}
    \hat{\psi}_{+,+}^{(1)} \\
    \hat{\psi}_{+,-}^{(1)} \\
    \hat{\psi}_{+,+}^{(2)} \\
    \hat{\psi}_{+,-}^{(2)} 
    \end{pmatrix}  = \frac{1}{\sqrt{2}} \begin{pmatrix}
        \widetilde{\psi}_{+}^{(1)} + \sigma_3  \widetilde{\psi}_{-}^{(1)} \\
        \widetilde{\psi}_{+}^{(1)} - \sigma_3  \widetilde{\psi}_{-}^{(1)} \\
        \widetilde{\psi}_{+}^{(2)} - \sigma_3  \widetilde{\psi}_{-}^{(2)} \\
        \widetilde{\psi}_{+}^{(2)} + \sigma_3  \widetilde{\psi}_{-}^{(2)}
    \end{pmatrix} \, , \notag\\
    \begin{pmatrix}
        \mathcal{U}_b \widetilde{\Psi}^{(1)} \\
        \mathcal{U}_a \widetilde{\Psi}^{(2)}
    \end{pmatrix}  &= \begin{pmatrix}
    \hat{\psi}_{-,+}^{(1)} \\
    \hat{\psi}_{-,-}^{(1)} \\
    \hat{\psi}_{-,+}^{(2)} \\
    \hat{\psi}_{-,-}^{(2)} 
    \end{pmatrix}= \frac{1}{\sqrt{2}} \begin{pmatrix}
        \widetilde{\psi}_{+}^{(1)} - \sigma_3  \widetilde{\psi}_{-}^{(1)} \\
        \widetilde{\psi}_{+}^{(1)} + \sigma_3  \widetilde{\psi}_{-}^{(1)} \\
        \widetilde{\psi}_{+}^{(2)} + \sigma_3  \widetilde{\psi}_{-}^{(2)} \\
        \widetilde{\psi}_{+}^{(2)} - \sigma_3  \widetilde{\psi}_{-}^{(2)}
    \end{pmatrix}  \, .
\end{align}
Using the fermions in Eq.~\eqref{eq:Psi-transformed-by-U}, the equations of motion in Eqs.~\eqref{eq:eom-for-mathcal-Pb-plus}--\eqref{eq:eom-for-mathcal-Pb-minus} become
\begin{align}
     \left[ \hat{\psi}^{(1)}_{-,-} (p,x+1) + m'  \hat{\psi}^{(1)}_{-,-} (p,x) \right] &=    0 \, , \\
     \left[ \hat{\psi}^{(2)}_{-,-} (p,x+1) + m'  \hat{\psi}^{(2)}_{-,-} (p,x) \right] &=    0 \, , \\
     \left[ \hat{\psi}^{(1)}_{+,-} (p,x-1) + m'  \hat{\psi}^{(1)}_{+,-} (p,x) \right] &=    0 \, , \\
    \left[ \hat{\psi}^{(2)}_{+,-} (p,x-1) + m'  \hat{\psi}^{(2)}_{+,-} (p,x) \right] &=    0 \, .
\end{align}
Here we define the fermions uniformly as
\begin{align}
    \label{eq:DW-fermion-plus}
    \hat{\Psi}_{+} &= \begin{pmatrix}
    \hat{\Psi}_{+}^{(1)} \\
    \hat{\Psi}_{+}^{(2)}
    \end{pmatrix} := \begin{pmatrix}
        \hat{\psi}_{+,-}^{(1)} \\
         \hat{\psi}_{+,-}^{(2)}
    \end{pmatrix} =\frac{1}{\sqrt{2}} \begin{pmatrix}
        \widetilde{\psi}_{+}^{(1)} - \sigma_3  \widetilde{\psi}_{-}^{(1)} \\
        \widetilde{\psi}_{+}^{(2)} + \sigma_3  \widetilde{\psi}_{-}^{(2)}
    \end{pmatrix}  \, , \\
     \label{eq:DW-fermion-minus}
    \hat{\Psi}_{-} &=  \begin{pmatrix}
    \hat{\Psi}_{-}^{(1)} \\
    \hat{\Psi}_{-}^{(2)}
    \end{pmatrix} := \begin{pmatrix}
        \hat{\psi}_{-,-}^{(1)} \\
         \hat{\psi}_{-,-}^{(2)}
    \end{pmatrix} = \frac{1}{\sqrt{2}} \begin{pmatrix}
        \widetilde{\psi}_{+}^{(1)} + \sigma_3  \widetilde{\psi}_{-}^{(1)} \\
        \widetilde{\psi}_{+}^{(2)} - \sigma_3  \widetilde{\psi}_{-}^{(2)}
    \end{pmatrix}  \, .
\end{align}
We note that each fermion $\hat{\Psi}_{\pm}^{(f)}$ is a two-component Dirac fermion, and each $\hat{\Psi}_{\pm}$ constitutes a two-flavor matrix fermion in $2+1$-dimensional spacetime.

For normalizability around $x \sim 0$, the fermions must satisfy
\begin{align}
    \hat{{\Psi}}_{+} : \, 
    \begin{cases}
        \vert m' \vert < 1  & \text{for}\,  x<0 \, , \\
         \vert m' \vert > 1  & \text{for}\,  x>0 \, , 
    \end{cases} \\
    \hat{{\Psi}}_{-} : \, 
    \begin{cases}
        \vert m' \vert > 1  & \text{for}\,  x<0 \, , \\
         \vert m' \vert < 1  & \text{for}\,  x>0 \, , 
    \end{cases} 
\end{align}
and around $x \sim \frac{N_x}{2}$,
\begin{align}
    \hat{{\Psi}}_{+} : \, 
    \begin{cases}
        \vert m' \vert < 1  & \text{for}\,  x<\frac{N_x}{2} \, , \\
         \vert m' \vert > 1  & \text{for}\,  x>\frac{N_x}{2} \, , 
    \end{cases} \\
    \hat{{\Psi}}_{-} : \, 
    \begin{cases}
        \vert m' \vert > 1  & \text{for}\,  x<\frac{N_x}{2} \, , \\
         \vert m' \vert < 1  & \text{for}\,  x>\frac{N_x}{2} \, , 
    \end{cases} 
\end{align}
it follows that the zero-mode $\hat{\Psi}_{-} (x)$ is localized on the domain wall at $x=0$, whereas the zero-mode $\hat{\Psi}_{+} (x)$ is localized on the domain wall at $x=\frac{N_x}{2}$, provided that
\begin{align}
    0 < am_0 < 3 \, .
\end{align}

\subsection{$2+1$~D systems on the Domain-wall}
\label{subsec:edge-systems-on-the-Domain-wall}

As discussed above, in the low-energy limit two-flavor free massless fermions are localized on the $2+1$~D domain walls at $x=0$ and $x=\frac{N_x}{2}$, respectively.
The corresponding 
low-energy effective Hamiltonians are given by
\begin{align}
    \label{eq:boundary-Hamiltonian-x-zero}
    H^{2D}_{(x=0)} &= \mathrm{i}  \sum_{\boldsymbol{R}} \hat{\Psi}_{-}^{\dagger} \left(\hat{\alpha}_i \otimes \mathbbm{1}\right) \partial_i \hat{\Psi}_{-} = \mathrm{i}  \sum_{\boldsymbol{R}}\sum_{f=1,2} \hat{\Psi}_{-}^{(f)\dagger}\hat{\alpha}_i \partial_i \hat{\Psi}_{-}^{(f)} \, , \\
    \label{eq:boundary-Hamiltonian-x-half}
    H^{2D}_{(x=\frac{N_x}{2})} &= \mathrm{i}  \sum_{\boldsymbol{R}} \hat{\Psi}_{+}^{\dagger} \left(\hat{\alpha}_i \otimes \mathbbm{1}\right) \partial_i \hat{\Psi}_{+} = \mathrm{i}  \sum_{\boldsymbol{R}}\sum_{f=1,2} \hat{\Psi}_{+}^{(f)\dagger}\hat{\alpha}_i \partial_i \hat{\Psi}_{+}^{(f)} \, , 
\end{align} 
where the index $i$ labels spatial directions, $i=1=y$ and $i=2=z$.
The $2\times2$ gamma matrices are
\begin{align}
    \hat{\beta}
    &= \hat{\gamma}_0 = \sigma_3   \,  , \, \hat{\alpha}_1 = \sigma_2 \, , \,  \hat{\alpha}_2 = -\sigma_1 \, , \, \\
    \hat{\gamma}_1 &= - \mathrm{i} \sigma_1 \, , \, \hat{\gamma}_2 = - \mathrm{i} \sigma_2 \, .
\end{align}

In this setup, recalling that the fermions $\hat{\Psi}_{\pm}$ given by Eqs.~\eqref{eq:tilde-Psi} and~\eqref{eq:Psi-transformed-by-U} are obtained via the unitary transformations in Eqs.~\eqref{eq:tilde-Psi} and~\eqref{eq:Psi-transformed-by-U}, we find that the discrete spacetime symmetries introduced in Sec.~\ref{subsec:discrete-symmetry} act on the domain-wall fermions as follows,
\begin{align}
    \label{eq:DW-parity-yz-action}
    \mathrm{P_{\perp}}:=\mathrm{R}_y\mathrm{R}_z&: \quad 
    \begin{pmatrix}
    \hat{\Psi}_{-} (x=0,t,\boldsymbol{R})\\
    \hat{\Psi}_{+} (x=\frac{N_x}{2},t,\boldsymbol{R})
    \end{pmatrix} \to (\hat{\beta} \otimes \sigma_3 \otimes \mathbbm{1})\, \begin{pmatrix}
    \hat{\Psi}_{-} (x=0,t,-\boldsymbol{R})\\
    \hat{\Psi}_{+} (x=\frac{N_x}{2},t,-\boldsymbol{R})
    \end{pmatrix} \, , \\ 
    \label{eq:DW-reflection-Y-action}
    \mathrm{R}_y&: \quad 
    \begin{pmatrix}
    \hat{\Psi}_{-} (x=0,t,\boldsymbol{R})\\
    \hat{\Psi}_{+} (x=\frac{N_x}{2},t,\boldsymbol{R})
    \end{pmatrix} \to ( \hat{\gamma}_1 \otimes \mathrm{i}\sigma_1 \otimes \sigma_3)\, \begin{pmatrix}
    \hat{\Psi}_{-} (x=0,t,\widetilde{\boldsymbol{R}_y}(\boldsymbol{R}))\\
    \hat{\Psi}_{+} (x=\frac{N_x}{2},t,\widetilde{\boldsymbol{R}_y}(\boldsymbol{R}))
    \end{pmatrix}, \\
    \label{eq:DW-reflection-Z-action}
    \mathrm{R}_z&: \quad 
    \begin{pmatrix}
    \hat{\Psi}_{-} (x=0,t,\boldsymbol{R})\\
    \hat{\Psi}_{+} (x=\frac{N_x}{2},t,\boldsymbol{R})
    \end{pmatrix} \to -( \hat{\gamma}_2 \otimes \mathrm{i}\sigma_2 \otimes \sigma_3)\, \begin{pmatrix}
    \hat{\Psi}_{-} (x=0,t,\widetilde{\boldsymbol{R}_z}(\boldsymbol{R}))\\
    \hat{\Psi}_{+} (x=\frac{N_x}{2},t,\widetilde{\boldsymbol{R}_z}(\boldsymbol{R}))
    \end{pmatrix}, \\
    \label{eq:DW-lattce-time-action}
    \mathrm{T}&: \quad 
     \begin{pmatrix}
    \hat{\Psi}_{-} (x=0,t,\boldsymbol{R})\\
    \hat{\Psi}_{+} (x=\frac{N_x}{2},t,\boldsymbol{R})
    \end{pmatrix} \to  (\hat{\gamma}_2 \otimes \sigma_2 \otimes \mathbbm{1}  )\, \begin{pmatrix}
    \hat{\Psi}_{-} (x=0,-t,\boldsymbol{R})\\
    \hat{\Psi}_{+} (x=\frac{N_x}{2},-t,\boldsymbol{R})
    \end{pmatrix}  \, , \\
     \label{eq:DW-lattice-charge-action}
    \mathrm{C}&: \quad 
     \begin{pmatrix}
    \hat{\Psi}_{-} (x=0,t,\boldsymbol{R})\\
    \hat{\Psi}_{+} (x=\frac{N_x}{2},t,\boldsymbol{R})
    \end{pmatrix} \to (\mathrm{i} \hat{\gamma}_1 \otimes \sigma_1 \otimes \mathbbm{1} )\, \begin{pmatrix}
    \hat{\Psi}_{-}^{*} (x=0,t,\boldsymbol{R})\\
    \hat{\Psi}_{+}^{*} (x=\frac{N_x}{2},t,\boldsymbol{R})
    \end{pmatrix} \, , \\
    \label{eq:DW-charge-action}
    \mathcal{C}&: \quad 
     \begin{pmatrix}
    \hat{\Psi}_{-} (x=0,t,\boldsymbol{R})\\
    \hat{\Psi}_{+} (x=\frac{N_x}{2},t,\boldsymbol{R})
    \end{pmatrix} \to (\hat{\gamma}_1 \otimes \sigma_3 \otimes \sigma_3 )\, \begin{pmatrix}
    \hat{\Psi}_{-}^{*} (x=0,t,\boldsymbol{R})\\
    \hat{\Psi}_{+}^{*} (x=\frac{N_x}{2},t,\boldsymbol{R})
    \end{pmatrix} \, , 
\end{align}
where the third matrices of the tensor act on edge-indices. 
Since the mass term $M_x^{(1)}$~\eqref{eq:Kink-mass} explicitly breaks the time-reversal symmetry generated by $\mathcal{T}$,
their actions exchange the two sides of the wall as
\begin{align}
    \label{eq:DW-time-action}
    \mathcal{T}&: \quad 
     \begin{pmatrix}
    \hat{\Psi}_{-} (x=0,t,\boldsymbol{R})\\
    \hat{\Psi}_{+} (x=\frac{N_x}{2},t,\boldsymbol{R})
    \end{pmatrix} \to (-\hat{\gamma}_2 \otimes \mathbbm{1} \otimes \sigma_1)\, \begin{pmatrix}
    \hat{\Psi}_{-} (x=0,-t,\boldsymbol{R})\\
    \hat{\Psi}_{+} (x=\frac{N_x}{2},-t,\boldsymbol{R})
    \end{pmatrix}  \, .
\end{align}
Therefore the symmetry cannot be considered as exact symmetry on the domain-wall.
Each edge mode~\eqref{eq:DW-fermion-plus}\eqref{eq:DW-fermion-minus} can be identified as an eigenstate of the reflection symmetry generated by $\mathrm{R}_x$, with eigenvalues $\pm 1$ since
\begin{align}
    \label{eq:DW-reflection-X-action}
    \mathrm{R}_x&: \quad 
    \begin{pmatrix}
    \hat{\Psi}_{-} (x=0,t,\boldsymbol{R})\\
    \hat{\Psi}_{+} (x=\frac{N_x}{2},t,\boldsymbol{R})
    \end{pmatrix} \to ( \mathbbm{1} \otimes \mathbbm{1} \otimes \sigma_3)\, \begin{pmatrix}
    \hat{\Psi}_{-} (x=0,t,\boldsymbol{R})\\
    \hat{\Psi}_{+} (x=-\frac{N_x}{2},t,\boldsymbol{R})
    \end{pmatrix} \, .
\end{align}

On each domain wall, the Dirac mass term $\hat{\Psi}_{\pm}^{\dagger}(\beta\otimes \mathbbm{1}) \hat{\Psi}_{\pm}$ is invariant under $\mathbb{Z}_2^{\mathrm{P}_{\perp}}$, $\mathbb{Z}_2^{\mathrm{C}}$, and $\mathbb{Z}_2^{\mathcal{C}}$, but not under $\mathbb{Z}_2^{\mathrm{R}_y}$, $\mathbb{Z}_2^{\mathrm{R}_z}$, or $\mathbb{Z}_2^{\mathrm{T}}$.
In contrast, the flavor mass term $\hat{\Psi}_{\pm}^{\dagger}(\beta \otimes \sigma_3 ) \hat{\Psi}_{\pm}$ is invariant under $\mathbb{Z}_2^{\mathrm{P}_{\perp}}$, $\mathbb{Z}_2^{\mathrm{R}_y}$, $\mathbb{Z}_2^{\mathrm{R}_z}$, $\mathbb{Z}_2^{\mathcal{C}}$, and $\mathbb{Z}_2^{\mathrm{T}}$, but not under $\mathbb{Z}_2^{\mathrm{C}}$.
Therefore $\mathrm{C}$--$\mathrm{R}_i$--$\mathrm{T}$ symmetries exclude all possible bilinear Dirac mass terms~\cite{Wan:2023nqe,Li:2024jle}.

It should be noted that $Q_{y}$ and $Q_{z}$, defined in Eq.~\eqref{eq:Qi-ConservedCharges-satisfying-OnsagerAlgebras} are conserved in the $2+1$~D boundary system, and these charges act on the fermions as
\begin{align}
    \label{eq:DW-Q2-transformation}
    \lim_{N \to \infty}[Q_{y}, \begin{pmatrix}
    \hat{\Psi}_{-} (x=0,t,\boldsymbol{R})\\
    \hat{\Psi}_{+} (x=\frac{N_x}{2},t,\boldsymbol{R})
    \end{pmatrix}] &= ( \mathbbm{1} \otimes \sigma_2 \otimes \mathbbm{1} ) \begin{pmatrix}
    \hat{\Psi}_{-} (x=0,t,\boldsymbol{R})\\
    \hat{\Psi}_{+} (x=\frac{N_x}{2},t,\boldsymbol{R})
    \end{pmatrix} \, , \\
    \label{eq:DW-Q3-transformation}
    \lim_{N \to \infty}[Q_{z}, \begin{pmatrix}
    \hat{\Psi}_{-} (x=0,t,\boldsymbol{R})\\
    \hat{\Psi}_{+} (x=\frac{N_x}{2},t,\boldsymbol{R})
    \end{pmatrix}] &= - ( \mathbbm{1} \otimes \sigma_1 \otimes \mathbbm{1} ) \begin{pmatrix}
    \hat{\Psi}_{-} (x=0,t,\boldsymbol{R})\\
    \hat{\Psi}_{+} (x=\frac{N_x}{2},t,\boldsymbol{R})
    \end{pmatrix} \, .
\end{align}
since the Hamiltonian~\eqref{eq:DW-Hamiltonian} possesses same symmetries with $\delta H_x ^{(1)}$.
Notably, $Q_y$ and $Q_z$ generate an $\mathrm{SU}(2)$ flavor symmetry on each domain wall, which we refer to as the valley $\mathrm{SU}(2)$ symmetry.
Ref.~\cite{Pace:2025rfu} demonstrated that conserved charges satisfying the Onsager algebra in a $2+1$~D honeycomb lattice system give rise to an $\mathrm{SU}(2)$ symmetry in the continuum limit, indicating that the valley symmetry is not emergent but ''emanant".
In the present construction, this emanant valley $\mathrm{SU}(2)$ symmetry originates from the conserved charges $Q_y$ and $Q_z$ defined in the $3+1$~D staggered fermion system.~\footnote{
We refer to such an IR symmetry, which arises from exact conserved lattice operators obeying a different algebra, as ``emanant'' \cite{Cheng:2022sgb,Seiberg:2023cdc,Barkeshli:2025cjs}. 
In our system, by ``emanant,'' we mean a boundary symmetry or anomaly that is already encoded in the bulk system, rather than one that merely emerges in the IR limit.}~\footnote{The commutator $[Q_y,Q_z]$, which is itself a bulk operator, likewise acts nontrivially on the zero modes and supplies the remaining $\mathrm{su}(2)$ generator; see Appendix~\ref{sec:app-extraEvidence}.} 

The valley $\mathrm{SU}(2)$ symmetry forbids Dirac flavor mass terms of the form $\hat{\Psi}_{\pm}^{\dagger}(\beta \otimes \sigma_i ) \hat{\Psi}_{\pm}$.
Consequently, on the domain wall there exists no mass term that is simultaneously invariant under the spacetime reflection symmetries $\mathbb{Z}_2^{\mathrm{R}_y}$, $\mathbb{Z}_2^{\mathrm{R}_z}$, and $\mathbb{Z}_2^{\mathrm{T}}$, as well as the $\mathrm{SU}(2)$ flavor symmetry.
This establishes the parity anomaly associated with the flavor $\mathrm{SU}(2)$ symmetry and space-reflection symmetry $\mathbb{Z}_2^{\mathrm{R}_i}$~\cite{Seiberg:2016rsg,Pace:2025rfu}.\footnote{Since the anomalous time-reversal symmetry $\mathbb{Z}_4^{\mathcal{T}}$ is explicitly broken while the time-reversal symmetry $\mathbb{Z}_2^{\mathrm{T}}$ is anomaly free~\cite{Pace:2025rfu} because it commutes with the $\mathrm{SU}(2)$ symmetry, the parity anomaly associated with time-reversal symmetry $\mathbb{Z}_4^{\mathcal{T}}$ is emergent anomaly.}
The anomaly associated with this emanant symmetry may be understood as a Lieb-Schultz-Mattis (LSM) anomaly~\cite{Lieb:1961fr, Affleck:1986pq, Oshikawa:2000zwq, Hastings:2003zx}; although we describe the $2+1$~D domain-wall theory in the continuum limit, the relevant symmetries are fundamentally exact on the underlying lattice.

Indeed, this anomaly can be trivialized in the following way which implies the anomaly is of order $2$.
Let us consider a Hamiltonian given by
\begin{align}
    H_{2DW} &= H_{DW\uparrow} + H_{DW\downarrow} \, ,
\end{align}
where $H_{DW\uparrow}$ is defined by Eq.~\eqref{eq:DW-Hamiltonian} and 
\begin{align}
     H_{DW\downarrow} = H - M_x ^{(1)} \, .
\end{align}
This Hamiltonian has the following zero modes, $\hat{\Psi}_{\uparrow,-}$ and $\hat{\Psi}_{\downarrow,+}$ analogous to the fermions~\eqref{eq:DW-fermion-minus} \eqref{eq:DW-fermion-plus} on the edge at $x=0$.
On the edge at $x=0$, the zero modes can form a mass as
\begin{align}
    \mathcal{M} = \sum_x \left[ \mathrm{i} \hat{\Psi}_{\uparrow,-}^{\dagger} \beta  \hat{\Psi}_{\downarrow,+} + \text{h.c.}  \right] \, ,
\end{align}
where $\beta$ acts on the spin indices.
This mass term commutes with conserved charges $Q_{\uparrow,y}+Q_{\downarrow,y}$ and $Q_{\uparrow,z}+Q_{\downarrow,z}$ associated with the Onsager algebra, and also respects the space-reflection symmetries generated by $\mathrm{R}_i$.
This result implies that the parity anomaly is of order $2$, which is consistent with~\cite{Seiberg:2016rsg,Pace:2025rfu}.

\section{Conclusion}
\label{sec:Conclusion}

The Hamiltonian of staggered fermion in $3+1$~D describes a massless free two-flavor Dirac fermion system. Because the spin and flavor degrees of freedom are staggered over the lattice, the Hamiltonian exhibits a rich symmetry structure.

In this paper, we have shown that, 
in addition to discrete spacetime symmetries, it possesses non-singlet $\mathrm{U}(1)_{F_i}$ subgroups of the axial flavor symmetry $\mathrm{SU}(2)_L \times \mathrm{SU}(2)_R \times \mathrm{U}(1)_A$, generated by conserved charges that satisfy the Onsager algebra.
We have further demonstrated that this system admits a wide variety of Dirac mass terms. Since a Dirac fermion is reconstructed from eight one-component staggered fermions residing within a unit cube, the admissible mass terms are not limited to on-site bilinears; they also arise from hopping terms extending over one-, two-, and three-link separations inside the cube.

Our systematic study on the classification of the Dirac mass terms is a central achievement of our work, as it intrinsically enables a far more detailed and rigorous investigation of 't~Hooft anomalies.
Among the classified possibilities, the one-link mass in the $x$-direction preserves a significantly larger residual symmetry group than the others. 
In particular, because the shift symmetries in the transverse directions remain intact, this mass term commutes with the conserved charges $Q_y$ and $Q_z$ that generate the Onsager algebra. 
We emphasize that this conclusion is robust: it universally holds for any staggered phase included in the ``normal definition,'' since all such phases are unitarily equivalent to the one given in Eq.~\eqref{eq:staggered-phase}.

Motivated by this large residual symmetry, we analyzed a Hamiltonian with a one-link mass in the $x$-direction with a kink profile along the $x$-direction. The resulting domain wall supports two flavors of two-component massless Dirac fermions localized on the wall. Remarkably, the bulk conserved charges $Q_y$ and $Q_z$, which generate the non-singlet $\mathrm{U}(1)_{F_i}$ symmetries in $3+1$~D, act as generators of a flavor $\mathrm{SU}(2)$ symmetry in the $2+1$~D boundary theory. 
Moreover, the commutators $G_{ij} := [Q_i, Q_j]$ with $i \neq j$ are themselves conserved charges of the lattice Hamiltonian, and in the continuum limit they generate, together with the $Q_i$, the flavor symmetry of the resulting theory. All the generators of the $\mathrm{SU}(2)_L \times \mathrm{SU}(2)_R$ symmetry of the $3+1$~D bulk theory, as well as those of the flavor $\mathrm{SU}(2)$ symmetry of the $2+1$~D boundary theory, are therefore realized by conserved charges that already exist on the lattice. These symmetries are thus emanant rather than emergent, which in turn underlies the anomaly structure discussed below. See Appendix~\ref{sec:app-extraEvidence} for further details.

Crucially, our comprehensive mass classification allows us to sharply delineate the resulting anomaly structure. Since the time-reversal symmetry $\mathbb{Z}_2^{\mathrm{T}}$ generated by $\mathrm{T}$ commutes with the $\mathrm{SU}(2)$ flavor symmetry, there is no parity anomaly between them. Consequently, in our system, the parity anomaly involving time-reversal and the flavor $\mathrm{SU}(2)$ symmetry is realized as an ``emergent" anomaly. On the other hand, the spatial reflection symmetry $\mathbb{Z}_2^{\mathrm{R}_i} \, (i=y,z)$ generated by $\mathrm{R}_i$ remains an exact symmetry in both the $3+1$~D bulk system and the $2+1$~D edge system. Therefore, the parity anomaly involving the exact spatial reflection and the $\mathrm{SU}(2)$ symmetry is identified as an ``emanant" anomaly.
This result implies that the emanant anomaly may be understood as a Lieb-Schultz-Mattis (LSM) anomaly; although we analyze the $2+1$~D edge theory in the continuum limit, the symmetries under consideration are fundamentally exact on the underlying lattice.
This clear distinction is a direct consequence of our precise symmetry analysis, and our result is partially consistent with the analysis of the $2+1$~D honeycomb lattice theory~\cite{Pace:2025rfu}.

This phenomenon can be naturally understood through the lens of anomaly inflow~\cite{Callan:1984sa}.\footnote{A quintessential lattice example of such bulk-boundary correspondence in Symmetry-Protected Topological (SPT) phases is the $1+1$~D cluster model~\cite{Raussendorf:2001, Chen:2011}. Protected by a $\mathbb{Z}_2 \times \mathbb{Z}_2$ symmetry, its fully gapped bulk dictates the inevitable appearance of degenerate zero-energy modes at the open boundaries. These localized modes transform projectively under the symmetry, which can be elegantly understood as a $0+1$~D quantum mechanical manifestation of anomaly inflow.
Furthermore, the rigorous mathematical foundation bridging domain-wall (or overlap) fermions, the Atiyah-Patodi-Singer (APS) index theorem, and anomaly inflow has been extensively developed by Refs.~\cite{Fukaya:2017tsq, Fukaya:2019zvj}.}
While the $3+1$~D closed bulk system is fully gapped and preserves both the exact spatial reflection symmetries and the continuous symmetries generated by $Q_y$ and $Q_z$, it realizes a non-trivial symmetry-protected topological (SPT) phase. Consequently, when a $2+1$~D boundary is introduced, anomaly inflow dictates the inevitable appearance of gapless edge modes. These boundary fermions must strictly carry the 't~Hooft anomalies corresponding to the bulk SPT order, thereby providing a robust physical mechanism for the emanant parity anomaly observed in our boundary theory.

Both in the intrinsic $2+1$~D theory and in the $2+1$~D theory obtained via dimensional reduction, the conserved charges forming the Onsager algebra generate a flavor $\mathrm{SU}(2)$ symmetry. This observation indicates that the symmetry realized by the Onsager algebra crucially depends on the number of spatial directions in which the shift symmetry is preserved.

Several directions for future investigation are suggested by our analysis.
First, it would be intriguing to construct an effective lattice Hamiltonian for the $2+1$~D boundary theory corresponding to Eqs.~\eqref{eq:boundary-Hamiltonian-x-zero} and~\eqref{eq:boundary-Hamiltonian-x-half}. Since our findings regarding the parity anomaly can be naturally understood in the context of anomaly inflow as we mentioned above, we expect the boundary lattice Hamiltonian to feature a Dirac operator analogous to that of overlap fermions~\cite{Ginsparg:1981bj,Luscher:1998pqa,Neuberger:1998wv}.
A natural question that arises is what analog of the Ginsparg-Wilson relation such a boundary Dirac operator would obey.
Second, it would be interesting to determine which symmetries are generated by the Onsager-algebra charges in a $4+1$~D Hamiltonian system, and how these charges behave in the $3+1$~D theory obtained via dimensional reduction. Third, it would be worthwhile to further explore the boundary setup constructed in this work using the deformation $\delta H_x^{(1)}$, which preserves two conserved charges satisfying the Onsager algebra. Since this deformation allows us to realize a $3+1$~D theory with a boundary while retaining comparatively large symmetries, one may investigate fermion scattering processes in this background. In particular, numerical studies of bulk fermion scattering in the presence of the kink-type mass profile may provide a useful toy model for fermion scattering off localized defects, potentially offering qualitative insight into chiral dynamics and fermion--monopole scattering phenomena~\cite{Razamat:2020kyf,Tong:2021phe,Xu:2025hfs}. It would be particularly fascinating if the conserved charges satisfying the Onsager algebra played a significant role in constraining such scattering processes.

\acknowledgments
The work of T.M. is supported in part by Japan Society for the Promotion of Science (JSPS) KAKENHI Grant No. 23K03425 and 22H05118.
The work of T.O. is supported in part by JSPS KAKENHI Grant Number 23K03387.
The work of T.Y. is supported in part by JST SPRING, Grant Number JP-MJSP2138.

\appendix

\section{Taste-splitting mass in Lagrangian formalism}
\label{sec:appendix_staggered_wilson}

While Section \ref{sec:classification-mass} focuses on the classification of mass terms for staggered fermions in the $3+1$~D Hamiltonian formalism, it is highly instructive to compare these structures with the $4$D Euclidean Lagrangian formalism \cite{Adams:2009eb, Adams:2010gx, Hoelbling:2010jw, Creutz:2011cd, deForcrand:2012bm, Misumi:2012sp}. 

\subsection{Taste-splitting mass}

Here we adopt the following notation:
$x$ stands for the four-dimensional coordinate $x=(x_1, x_2 , x_3 , x_4)$
and $\chi_{x}$ is a one-component staggered field at the lattice site $x$. 
We also define $\xi_\mu \equiv \gamma^{T}_\mu$ as gamma matrices in the taste space.

Before introducing the taste-splitting mass,
we review the free $4$D staggered action,
\begin{equation}
    S = \sum_{x,y} \bar{\chi}_x \left[ \eta_\mu D_\mu  + M \right]_{xy} \chi_y,
\end{equation}
where $(\eta_{\mu})_{xy}=\eta_{\mu}(x)\delta_{x,y}$ with $\eta_{\mu}(x) \equiv (-1)^{x_{1}+...+x_{\mu-1}}$, $D_\mu = (V_\mu - V_\mu^\dagger)/2$ with $(V_{\mu})_{xy} = \delta_{y,x+\mu}$ and $M = m_0\, \delta_{x,y}$ is an on-site mass. 
For $m_0 =0$, the action describes four Dirac species (tastes) while preserving the staggered chiral symmetry 
$\chi_{x}\,\to\, e^{{\rm i} \theta\epsilon(x)}\chi_{x}$, $\bar{\chi}_{x}\,\to\, \bar{\chi}_{x} e^{{\rm i} \theta\epsilon(x)}$  with $\epsilon(x) = (-1)^{x_1+x_2+x_3+x_4} $.

In this $4$D Lagrangian formalism of staggered fermions, unlike the case of the Hamiltonian formalism in Sec.~\ref{sec:classification-mass}, there are only two types of taste-splitting mass terms satisfying $\gamma_5$-hermiticity, which is essential for the positivity of fermion determinants \cite{Adams:2009eb, Hoelbling:2010jw}. 
It should be stressed that the Hamiltonian formalism and the Euclidean Lagrangian formalism are different setups subject to different constraints: Hermiticity in the former and $\gamma_5$-hermiticity in the latter. Therefore, the allowed taste-splitting mass terms need not coincide.

The first type is the Adams-type mass $M_A$, which is a four-link term \cite{Adams:2009eb, Adams:2010gx},
\begin{equation}  
M_{A} = \epsilon\sum_{sym} \eta_{1}\eta_{2}\eta_{3}\eta_{4}
C_{1}C_{2}C_{3}C_{4}
= (\mathbf{1} \otimes \xi_{5}) + O(a),
\label{AdamsM}
\end{equation}
where $C_\mu = (V_\mu + V_\mu^\dagger)/2$,  $(\epsilon)_{xy}=\epsilon(x)\delta_{x,y}$ and
the factor $1/24$ is hidden in the symmetric sum $\sum_{sym}$.
The spin-taste representation on the right-hand side means that
the four tastes split into two tastes in the $\xi_{5}=+1$ subspace and two tastes in the $\xi_{5}=-1$ subspace ($2+2$ splitting).
With a proper mass shift, one obtains the Adams-type staggered-Wilson fermion,
\begin{equation}
S_A = \sum_{x,y} \bar{\chi}_x \left[ \eta_\mu D_\mu + r(1 + M_A) + M \right]_{xy} \chi_y,
\end{equation}
where $r$ is the Wilson parameter.

The second type is the Hoelbling-type mass $M_H$ \cite{Hoelbling:2010jw}, given by a two-link term,
\begin{align}
M_H &= {\rm i} ( \eta_{\mu\nu} C_{\mu\nu} + \eta_{\rho\sigma} C_{\rho\sigma})
\nonumber \\
    &= [\mathbf{1} \otimes (\sigma_{\mu\nu} + \sigma_{\rho\sigma})] + \mathcal{O}(a),
\end{align}
where $(\eta_{\mu\nu})_{xy} \equiv \epsilon_{\mu\nu}(x) \eta_{\mu}(x) \eta_{\nu}(x) \delta_{x,y}$, $\epsilon_{\mu\nu}(x) \equiv (-1)^{x_\mu+x_\nu}$, and $C_{\mu\nu} \equiv (C_\mu C_\nu + C_\nu C_\mu)/2$. 
Here, $\mu,\nu,\rho,\sigma$ are permutations of $1,2,3,4$ and we define $\sigma_{\mu\nu} \equiv {\rm i} \xi_{\mu}\xi_{\nu} $.
Since the eigenvalues of $\sigma_{\mu\nu} + \sigma_{\rho\sigma}$ are $0,\,\pm2$, this taste-splitting mass splits four tastes into three branches: one with positive mass, two with zero mass, and one with negative mass ($1+2+1$ splitting).
The Hoelbling-type staggered-Wilson fermion is given by
\begin{equation}
S_H = \sum_{x,y} \bar{\chi}_x \left[ \eta_\mu D_\mu + r(2 + M_H) + M \right]_{xy} \chi_y.
\end{equation}
All the other possibilities, including one-link and three-link mass terms, are ruled out by the $\gamma_5$-hermiticity.

\subsection{Discrete Symmetries}

The two types of taste-splitting mass terms for $4$D staggered fermions break some discrete staggered symmetries as well as the staggered chiral symmetry \cite{Misumi:2012sp,Misumi:2012eh,Misumi:2020eyx,Chreim:2024duv}.
We begin with a review of the original staggered symmetry (apart from $U(1)_V$) \cite{Golterman:1984cy}, which is expressed as
\begin{equation}
\{ C_{0}, \,S_{\mu},\, I_{S},\, R_{\mu\nu} \}\,\,\times\,\,\{U^{\epsilon}\}_{m_0 =0}.
\label{sSym}
\end{equation}
Here $C_{0}$ is the staggered charge conjugation, 
\begin{equation}
    C_0 : \chi_x \to \epsilon(x) \bar{\chi}_x, \quad \bar{\chi}_x \to -\epsilon(x) \chi_x .
\end{equation}
$S_{\mu}$ is the shift transformation, 
\begin{equation}
S_\mu : \chi_x \to \zeta_\mu(x)\chi_{x+\hat{\mu}}, \quad \bar{\chi}_x \to \zeta_\mu(x)\bar{\chi}_{x+\hat{\mu}},
\end{equation}
with $\zeta_1(x)=(-1)^{x_2+x_3+x_4}$, $\zeta_2(x)=(-1)^{x_3+x_4}$, $\zeta_3(x)=(-1)^{x_4}$, and $\zeta_4(x)=1$.
$I_{S}$ is the spatial inversion, 
\begin{equation}
I_S : \chi_x \to \eta_{4}(x) \chi_{I_S^{-1} x},
\end{equation}
where under the spatial inversion $I_S$ the space coordinates transform as $x_j \to -x_j$ for $j=1,2,3$. 
$R_{\mu\nu}$ is the hypercubic rotation, 
\begin{align}
R_{\mu\nu}:\,\,\chi_{x} \to S_{R}(R^{-1}x)\chi_{R^{-1}x},\,\,\,\,
\bar{\chi_{x}} \to S_{R}(R^{-1}x)\bar{\chi}_{R^{-1}x},
\label{rot1}
\end{align}
where
$S_{R}(x)={1\over{2}}[1\pm\eta_{\mu}(x)\eta_{\nu}(x)\mp\zeta_{\mu}(x)\zeta_{\nu}(x)
+\eta_{\mu}(x)\eta_{\nu}(x)\zeta_{\mu}(x)\zeta_{\nu}(x)]$ with 
$\rho$\hspace{0.3em}\raisebox{0.4ex}{$<$}\hspace{-0.75em
}\raisebox{-0.7ex}{$>$}\hspace{0.3em}$\sigma$ and under the rotation $R_{\mu\nu}$ the spacetime indices transform as $\mu\to\nu,\,\nu\to -\mu$.
$U^{\epsilon}$ is the staggered chiral symmetry.

Both taste-splitting mass terms break the shift $S_\mu$ and the spatial inversion $I_S$, while both are invariant under the combined transformation $S_\mu I_S $, which leads to the physical parity in the continuum limit.

The Adams-type mass is invariant under the staggered charge conjugation $C_0$ while the Hoelbling-type mass breaks it. However, the following modified charge conjugation survives for the Hoelbling-type mass,
\begin{equation}
C_T = R_{\mu\rho}^2 C_0 ,
\end{equation}
which also serves as the physical charge conjugation in the continuum limit. We note that $R_{\nu\sigma}^2 C_0$ is also symmetry for the Hoelbling-type mass.

Regarding the rotational symmetry, the Adams-type mass preserves the full rotational symmetry $R_{\mu\nu}$.
The combination of the rotational symmetry $R_{\mu\nu}$ and $S_\mu I_S $ ensures the recovery of full Euclidean Lorentz symmetry. Conversely, the Hoelbling-type mass explicitly breaks the rotational symmetry into a ``doubled rotation'' subgroup, $R_{\rho\sigma}R_{\mu\nu}$ with $\mu,\nu,\rho,\sigma$ being arbitrary permutations of $1,2,3,4$ for the fixed Hoelbling mass. The loss of rotational symmetry implies that the full Euclidean Lorentz symmetry may not be recovered in the continuum limit.
This issue has recently been investigated in Ref.~\cite{Chreim:2024duv}. 

For the case without the on-site mass shift ($m_0 = 0$), the special charge conjugation 
\begin{equation}
C_T ' : \chi_x \to \bar{\chi}_x, \quad \bar{\chi}_x \to  \chi_x ,
\end{equation}
becomes relevant, which is given as the combination of $C_0$, $U^{\epsilon}$ staggered-chiral and $U(1)_V$ transformations.
For the Adams-type mass without the on-site mass shift, the combined symmetry $S_\mu I_S $ is enhanced to $C_{T}' S_\mu$ and $C_{T}' I_S$ although $C_{T}'$ itself is not symmetry.
In contrast, for the Hoelbling-type mass without the on-site mass shift, $C_{T}'$ is exact symmetry, which is a typical example of the {\it central-branch} symmetry \cite{Creutz:2011cd,Kimura:2011ik,Misumi:2012eh,Misumi:2019jrt,Misumi:2020eyx}.

To sum up,
the staggered symmetry of the Adams-type mass is
\begin{equation}
\{ C_{0}, \,S_{\mu}I_{S},\, R_{\mu\nu} \}_{m_0 \not=0}\,,\qquad
\{ C_{0}, \,C_{T}' S_\mu,\,C_{T}' I_S,\, R_{\mu\nu} \}_{m_0 =0}\,,
\end{equation}
while the staggered symmetry of the Hoelbling-type mass is
\begin{equation}
\{ C_{T}, \,S_{\mu}I_{S},\, R_{\mu\nu}R_{\rho\sigma} \}_{m_0 \not=0}\,,\qquad
\{ C_{T}, \,C_{T}' ,\,S_{\mu}I_{S},\, R_{\mu\nu}R_{\rho\sigma} \}_{m_0 =0}\,.
\end{equation}
These symmetries have intensively been investigated in Refs.~\cite{Misumi:2012sp,Misumi:2012eh,Misumi:2020eyx,Chreim:2024duv}.

\section{Extra evidence of emanant $\mathrm{SU}(2)$ symmetry on the edge}
\label{sec:app-extraEvidence}
In this section we provide additional evidence that the flavor $\mathrm{SU}(2)$ symmetry realized on the domain-wall is emanant, i.e., that it originates from conserved charges defined in the bulk theory. To this end, we show that not only the conserved charges $Q_y$ and $Q_z$ themselves, but also their bulk commutator, act nontrivially on the zero-mode subspace and reproduce the expected $\mathrm{su}(2)$ algebra.

Let us define the commutator of the conserved charges $Q_y$ and $Q_z$, given in Eq.~\eqref{subsec:Onsager-ConservedCharge}, as
\begin{align}
\label{eq:Gyz}
   G_{yz} &:= [Q_y, Q_z ] \notag \\
   &= \frac{1}{2} \sum_{\boldsymbol{r}} \zeta_y (\boldsymbol{r}) \left( a (\boldsymbol{r}) a (\boldsymbol{r} + \hat{y} - \hat{z}  ) +   b (\boldsymbol{r}) b (\boldsymbol{r} - \hat{y} + \hat{z}  )   \right) \, .
\end{align}
Although $G_{yz}$ is constructed entirely from bulk operators, its action on the one-component staggered fermion $\chi(\boldsymbol{r})$,
\begin{align}
    [G_{yz}, \chi(\boldsymbol{r})] &= - (-1)^z \left( \chi(\boldsymbol{r} -\hat{y} + \hat{z} ) +   \chi(\boldsymbol{r} +\hat{y} - \hat{z} )   \right) \, ,
\end{align}
shows that $G_{yz}$ generates a translation along the diagonal direction in the $y$-$z$ plane.

We now examine this action on the matrix fermion $\Psi(\boldsymbol{k})$ reconstructed in the continuum limit around the momenta $\boldsymbol{K}_0 = (0,0,0)$ and $\boldsymbol{K}_1 = (0,0,\pi)$. In the basis consistent with the flavor convention adopted in this paper, $\Psi(\boldsymbol{k})$ transforms as
\begin{align}
    \lim_{N \to \infty} [G_{yz}, \Psi(\boldsymbol{k})] &= 2 \mathrm{i} \left( \mathbbm{1}_{4\times4}  \otimes \sigma_1 ^T \right)  \Psi(\boldsymbol{k}) \, ,
\end{align}
where $\mathbbm{1}_{4\times4}$ and $\sigma_1$ act on the Dirac spinor space and the flavor space, respectively. Applying the basis transformation of Eq.~\eqref{eq:unitary-matrix-Lambda}, under which the flavor generator transforms as $\sigma_1^T \to \sigma_3^T$, we obtain
\begin{align}
    \lim_{N \to \infty} [G_{yz}, \widetilde{\Psi}(\boldsymbol{k})] &= 2 \mathrm{i} \left( \mathbbm{1}_{4\times4}  \otimes \sigma_3 ^T \right)   \widetilde{\Psi}(\boldsymbol{k}) \, ,
\end{align}
where $\widetilde{\Psi} = \widetilde{\Lambda} \Psi$ is defined in Eq.~\eqref{eq:tilde-Psi}. This is precisely the basis used in the construction of the domain-wall zero modes.

Finally, projecting $\widetilde{\Psi}$ onto the domain-wall localized modes $\widehat{\Psi}_{-}$ and $\widehat{\Psi}_{+}$, we find
\begin{align}
   \lim_{N \to \infty} [G_{yz} ,\begin{pmatrix}
    \hat{\Psi}_{-} (x=0,t,\boldsymbol{R})\\
    \hat{\Psi}_{+} (x=\frac{N_x}{2},t,\boldsymbol{R})
    \end{pmatrix} ] & =  2 \mathrm{i} ( \mathbbm{1} \otimes \sigma_3 \otimes \mathbbm{1} ) \begin{pmatrix}
    \hat{\Psi}_{-} (x=0,t,\boldsymbol{R})\\
    \hat{\Psi}_{+} (x=\frac{N_x}{2},t,\boldsymbol{R})
    \end{pmatrix} \, ,
\end{align}
where, as before, the first, second, and third matrix factors act on the $(2+1)$-dimensional Dirac spinor space, the flavor space, and the two-edge space, respectively.

Together with the results for $Q_y$ and $Q_z$ themselves, this shows that not only the bulk conserved charges but also their commutator act nontrivially on the domain-wall zero-modes and reproduce the $\mathrm{su}(2)$ algebra generators expected on the edge. All the generators of the $\mathrm{SU}(2)$ flavor symmetry on the edge therefore exist already at the level of the lattice bulk theory. This constitutes further evidence that the flavor $\mathrm{SU}(2)$ symmetry realized on the domain-wall is not an emergent symmetry of the edge theory, but rather an ``emanant'' symmetry inherited from the conserved charges of the lattice bulk Onsager algebra.

The computation for $G_{yz}$ above generalizes immediately to the remaining pairs of directions: quite generally,
\begin{align}
    G_{ij} := [Q_i, Q_j] \, , \qquad i \neq j \, ,
\end{align}
becomes, in the continuum limit, a generator of the vector flavor $\mathrm{SU}(2)$ symmetry in the bulk. This observation, together with the results obtained above, carries two further implications.

\paragraph{(i) $\mathrm{Ons}\,3$ fully generates $\mathrm{SU}(2)_L \times \mathrm{SU}(2)_R$ in the bulk.}
In the continuum limit, $Q_i$ and $G_{ij}$ are respectively the generators of $\mathrm{U}(1)_{F_i} \subset \mathrm{SU}(2)_L \times \mathrm{SU}(2)_R$ and of the vector $\mathrm{SU}(2)$ symmetry. Hence the Onsager algebra $\mathrm{Ons}\,3$, built entirely out of bulk lattice operators, generates the full chiral flavor symmetry $\mathrm{SU}(2)_L \times \mathrm{SU}(2)_R$ in the continuum limit. This gives further evidence that $\mathrm{SU}(2)_L \times \mathrm{SU}(2)_R$ is not an emergent symmetry of the continuum theory, but an emanant one, already realized exactly at the level of the lattice bulk theory.

\paragraph{(ii) Explicit breaking of $\mathrm{SU}(2)_L \times \mathrm{SU}(2)_R$ and the origin of the edge $\mathrm{SU}(2)$.}
As shown in Sec.~\ref{sec:classification-mass}, gapping the bulk fermion requires breaking at least one conserved charge $Q_i$, which from the continuum-theory viewpoint amounts to an explicit breaking of $\mathrm{SU}(2)_L \times \mathrm{SU}(2)_R$. In particular, the kink-profile one-link mass term along the $x$-direction, Eq.~\eqref{eq:Kink-mass}, employed in Sec.~\ref{sec:mass-splitting}, breaks $\mathrm{Ons}\,3$ while preserving only $Q_y$, $Q_z$, and $G_{yz}$. These three charges close onto $\mathrm{Ons}\,2$, which is precisely the $\mathrm{su}(2)$ algebra realized on the edge.

In summary, $\mathrm{Ons}\,3$ generates the chiral flavor symmetry $\mathrm{SU}(2)_L \times \mathrm{SU}(2)_R$ in the continuum limit and thereby forbids a fermion mass term, while $\mathrm{Ons}\,2$ -- the subalgebra surviving once $\mathrm{Ons}\,3$ is explicitly broken by such a mass term -- generates the flavor $\mathrm{SU}(2)$ symmetry realized on the edge.

\newpage
\bibliographystyle{utphys}
\bibliography{reference}

\end{document}